\documentclass{article}

\usepackage[dvips]{graphicx}
\usepackage{psfrag}
\usepackage{amssymb,latexsym}

\newcommand{\mylab}[3]{\raisebox{#2}[0mm][0mm]{%
\makebox[0mm][l]{\hspace*{#1}\textbf{#3}}}}
\def\spacce#1{\hskip #1pt}
\def\drawline#1#2{\raise 2.5pt\vbox{\hrule width #1pt height #2pt}}
\def\solid{\drawline{24}{.5}\nobreak\ }

\def\bdash{\hbox{\drawline{1}{.5}\spacce{2}}}

\def\dashed{\bdash\bdash\bdash\bdash\nobreak\ }

\def\bdot{\hbox{\drawline{1}{.5}\spacce{2}}}

\def\dotted{\hbox{\leaders\bdot\hskip 24pt}\nobreak\ }

\def\circle{$\circ$\nobreak }
\def\trian{\raise 1.25pt\hbox{$\scriptstyle\triangle$}\nobreak\ }
\def\dtrian{\raise 1.25pt\hbox%
{$\scriptscriptstyle\bigtriangledown$}\nobreak\ }
\def\squar{\raise 1.25pt\hbox{$\scriptstyle\Box$}\nobreak\ }
\def\diamon{\raise 1.25pt\hbox{$\scriptstyle\diamond$}\nobreak\ }
\def\solidtrian{$\blacktriangle$\nobreak\ }
\def\solidsquar{$\blacksquare$\nobreak\ }

\def\dd{\, \rm{d}}
\def\bra{\langle}
\def\ket{\rangle}

\def\KMM87{Kim, Moin \& Moser, 1987}
\def\MKM99{Moser, Kim \& Mansour, 1999}
\def\AKM01{Abe, Kawamura \& Matsuo, 2001}
\def\GE00{DeGraaf \& Eaton, 2000}
\def\SM83{Smith \& Metzler, 1983}
\def\JFM01{Jim\'enez, Flores \& Garc\'\i a-Villalba, 2001}
\def\MK01{Metzger \& Klewicki, 2001}
\def\PL90{Perry \& Li, 1990}
\def\PHC86{Perry, Henbest \& Chong, 1986}

\title{Direct numerical simulation of the very large anisotropic scales in
a turbulent channel}

\author{Juan C. del \'Alamo%
  \footnote{School of Aeronautics UPM, 28040 Madrid, Spain}
  \and Javier Jim\'enez%
  \footnote{Also at School of Aeronautics UPM, 28040 Madrid, Spain}
  }

\begin{document}
\maketitle

\section{Introduction}

\label{juan1}

Over the last decades the knowledge on the small scales of turbulent wall
flows has experienced a significant advance, especially in the near-wall
region where the highest production of turbulent energy and the maximum
turbulence intensity occur.  The development of computers has played an
important role in this progress, making direct numerical simulations
affordable (\KMM87), and offering wider observational possibilities than
most laboratory experiments.

The large scales have received less attention, and it has not been until
recently that their significance and their real size have been widely
recognized, thanks in part to the experiments by \cite{Hit97} and
\cite{KimAdr99}, and to the compilation of experimental and numerical data
by \cite{Jim98}.  Two are the main reasons for this.  In the first place,
when \cite{Tow76} originally proposed the existence of very large anisotropic
scales (VLAS) in the overlap layer under the `attached eddy' hypothesis, he
described them as `inactive', not containing Reynolds stresses.
\cite{PerHenCho86} repeated that assertion in their elaboration of
Townsend's model, and this has probably contributed to their relative
neglect by later investigators.  \cite{Jim98} showed however that this
characterization is only partly correct, and that the VLAS carry a
substantial fraction of the Reynolds stresses.  We will provide in this
report further evidence that they carry a substantial part of the turbulent
energy in the flow and that they are `active' in Townsend's sense.

The large size of these scales also makes them difficult to study, both
experimentally and numerically.  Many of the high-Reynolds number
laboratory experiments lack spectral information, have too few wall
distances, or have data records which are too short to capture the largest
scales.  Moreover, most of of them contain only streamwise information, and
data on the spanwise scales are scarce.  The requirements of both a very
large box and a high Reynolds number has made direct numerical simulation
of the VLAS unapproachable until today.  Previously available numerical
databases were restricted to low Reynolds numbers, with little or no
separation between the small and large scales (\KMM87), or to small
computational boxes which interfere with the VLAS (\MKM99; \AKM01).

The purpose of this report is to serve as a preliminary description of a
newly compiled numerical database of the characteristics of the large
scales in turbulent channel flow at moderate Reynolds numbers.

\section{The numerical experiment}

Our investigation has been carried on a direct numerical simulation of the
turbulent incompressible flow in plane channels at Reynolds numbers
$Re_\tau=180$ and $Re_\tau=550$, based on the wall friction velocity,
$u_\tau$, and on the channel half-width $h$.  The emphasis in this report
will be on the latter of those two simulations.  The numerical code is
fully spectral, using dealiased Fourier expansions in the streamwise and
spanwise directions, and Chebychev polynomials in the wall-normall one, as
in \cite{KimMoiMos87}.  Although there are computations in the literature
at somewhat higher, although comparable, Reynolds numbers (\MKM99; \AKM01;
see table \ref{tab:otherguys}), we believe that this is the first
simulation in which the numerical box is large enough not to interfere with
the largest structures in the flow.

\begin{table}
\begin{center}
\begin{tabular}{ccccccccccc}
\vspace{0cm} & $Re_{\tau}$ & ${\Delta x^+}$  & ${\Delta z^+}$ &
$\Delta y_{max}^+$ & $ L_x / h $ &
$L_z / h$ & $N_x$ & $N_z$ & $N_y$ & Numerics \\[1ex]
Moser {\it et al.} (1999) & $590$ & $7.2$ & $3.6$ & $7.2$ & $2 \pi$ & $\pi$
& $512$ & $512$ & $257$ & Spectral\\
Abe {\it et al.} (2001) & $640$ & $8.0$ & $5.0$ & $8.2$ & $6.4$ & $2$
& $512$ & $256$ & $256$ & Second order FD\\
Present case 1 & $550$ & $8.9$ & $4.5$ & $6.7$ & $8 \pi$ & $4 \pi$
& $1536$ & $1536$ & $257$ & Spectral\\
Present case 2 & $180$ & $8.9$ & $4.5$ & $6.1$ & $12 \pi$ & $4 \pi$
& $768$ & $512$ & $97$ & Spectral\\
\end{tabular}
\end{center}
\caption{Summary of cases.  The resolution is measured in collocation
points}
\label{tab:otherguys}
\end{table}

The experimental results for high-Reynolds number turbulent wall flows (see
the references given in the previous section) reveal that the premultiplied
one-dimensional streamwise velocity spectrum $k_x E_{uu}^{1D}(k_x,y)$ has
two peaks.  The first one is in the wall region and scales in wall units.
Its position does not vary with the distance $y$ to the wall and
corresponds to the size of the buffer layer streaks.  At the top of the
buffer layer, the first peak coexists with a second one which scales in
outer units and is characteristic of the outer region.  The second peak
becomes stronger as the Reynolds number increases and its position
corresponds to the VLAS.  Its length increases with $y$ and reaches a
maximum of $4 - 15\,\delta$ (where $\delta$ is the characteristic flow
thickness) at a wall distance which scales in outer units and which depends
on the type of flow.  Beyond that level, the peak moves to shorter
wavelengths, until the streamwise turbulent energy becomes associated to scales of
length $\lambda_x \approx \delta$ at $y = \delta$.  With this information
in mind, and with the aid of tests cases performed at $Re_\tau = 180$ and
$Re_\tau=550$ in boxes of different sizes, we have used a box of size $L_x
\times L_y \times L_z = 8\pi h \times 2h \times 4\pi h$ in the streamwise,
wall-normal and spanwise directions for our $Re_\tau=550$ simulation.

\begin{figure}
  \begin{minipage}[t]{0.48\textwidth}

\includegraphics[height=55mm,width=\textwidth]{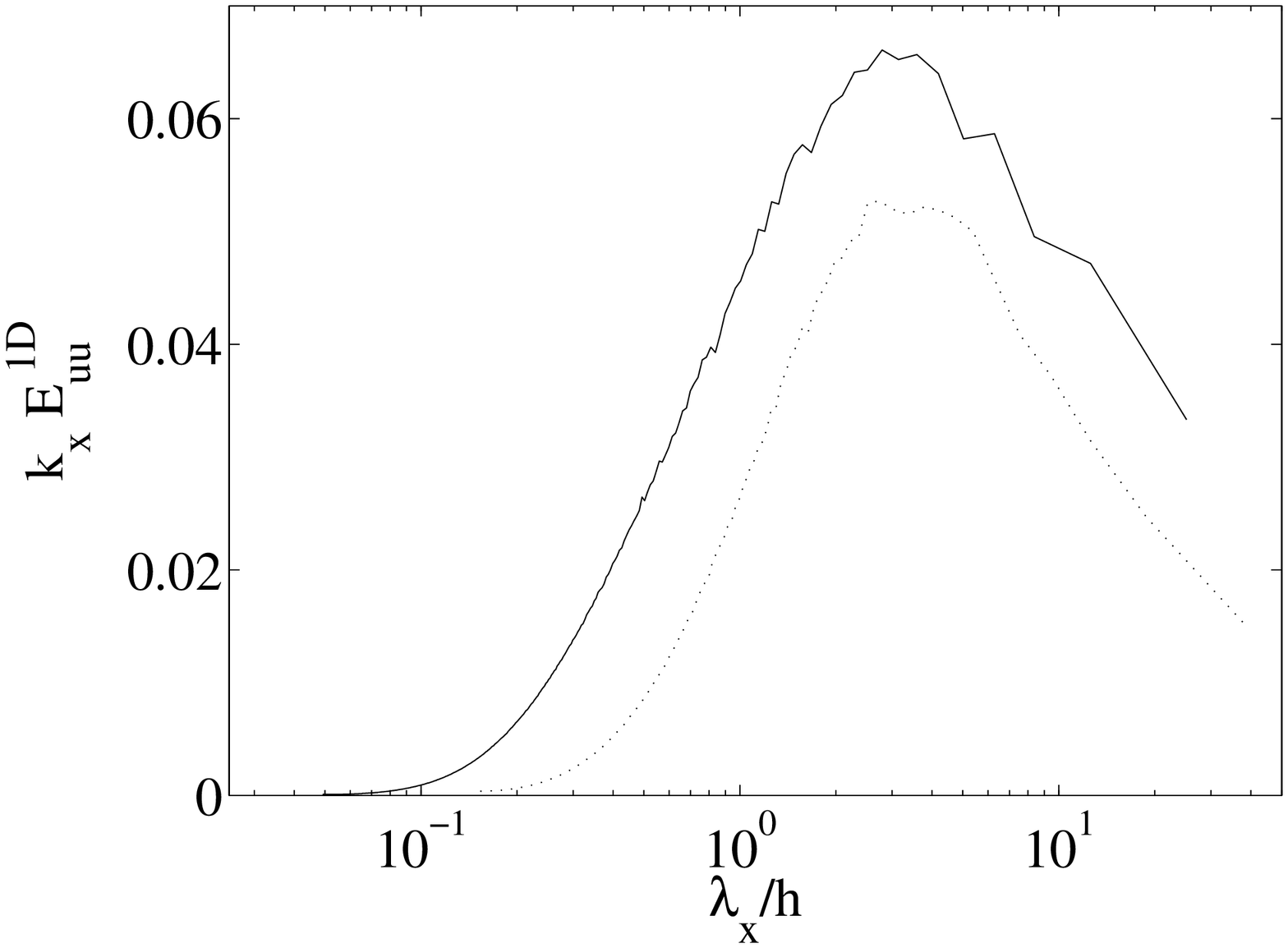}
     \mylab{.2\textwidth}{.8\textwidth}{\it (a)}%
   \end{minipage}
   \hfill
   \begin{minipage}[t]{0.48\textwidth}

\includegraphics[height=55mm,width=\textwidth]{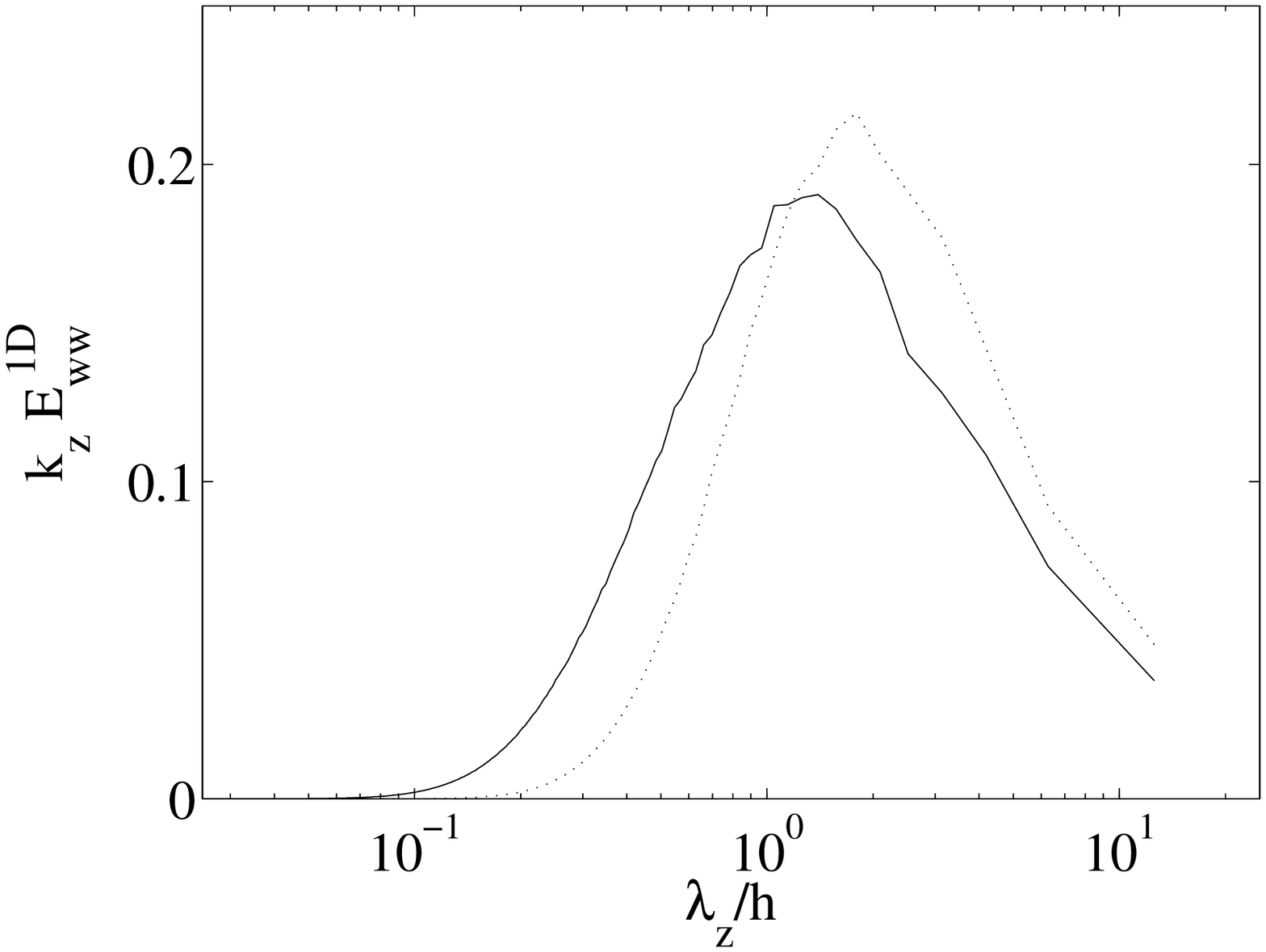}
      \mylab{.2\textwidth}{.8\textwidth}{\it (b)}%
   \end{minipage}
\caption{Premultiplied one-dimensional spectra at  $y=0.5h$. \solid,
    present $Re_\tau=550$; \dotted, present $Re_\tau=180$.
   {\it (a)} $k_xE_{uu}^{1D}(k_x)$. {\it (b)} $k_zE_{ww}^{1D}(k_z)$;
}
   \label{fig:sp1d}
\vspace{3mm}
   \begin{minipage}[t]{0.48\textwidth}
      \includegraphics[height=50mm,width=\textwidth]{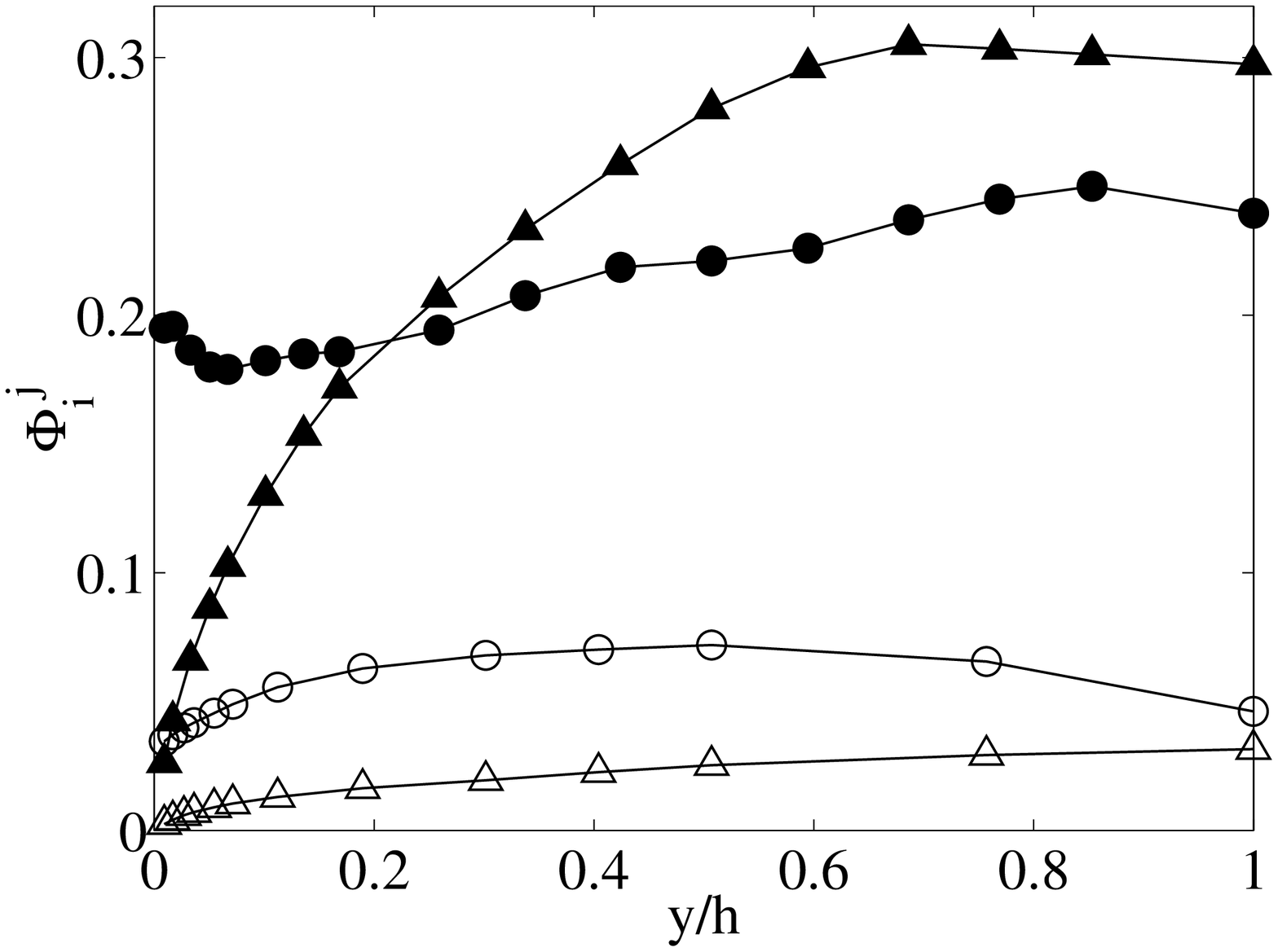}
      \mylab{.2\textwidth}{47mm}{\it (a)}%
   \end{minipage}
   \hfill
   \begin{minipage}[t]{0.48\textwidth}

\includegraphics[height=50mm,width=\textwidth]{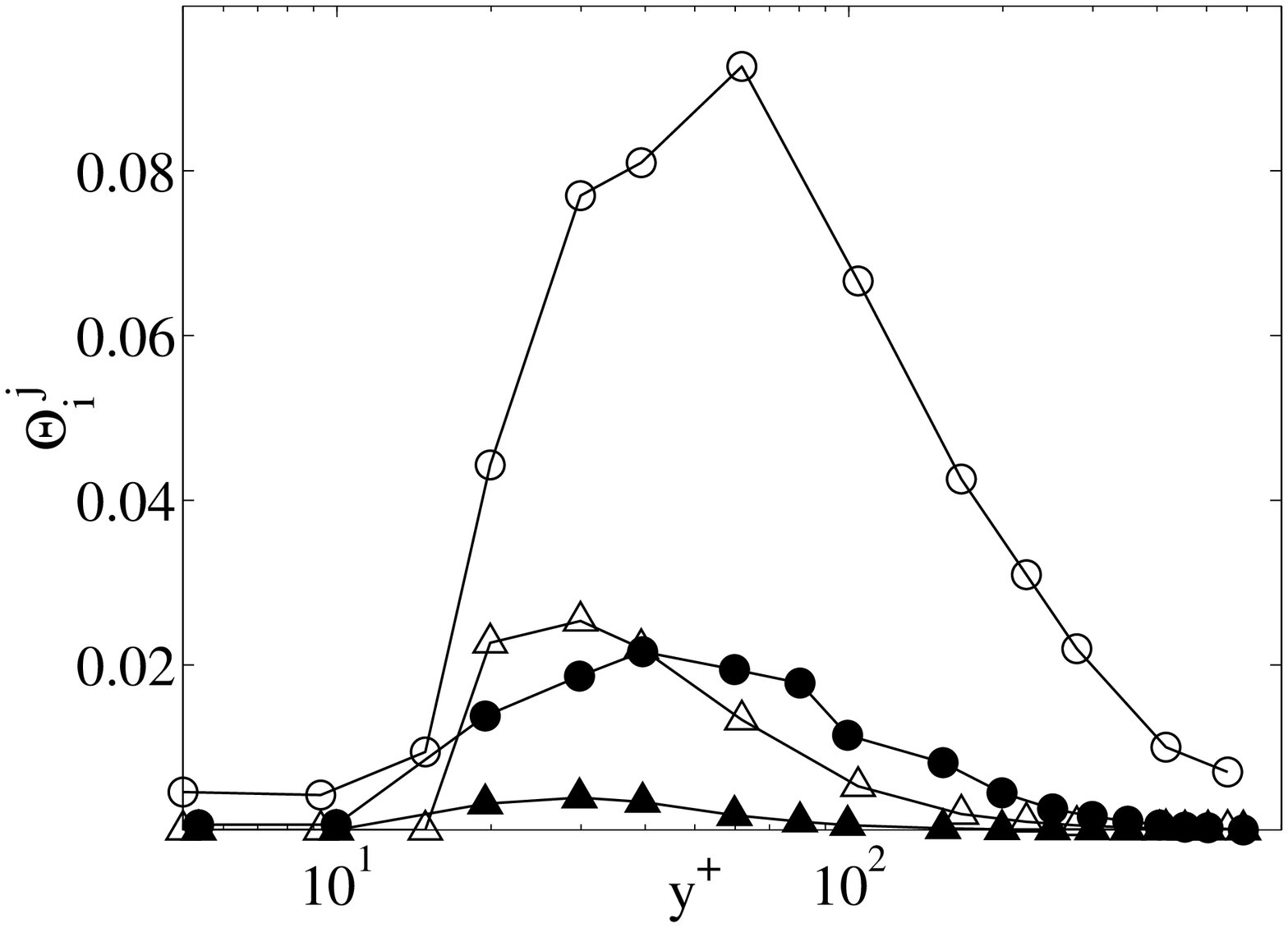}
      \mylab{.2\textwidth}{47mm}{\it (b)}%
   \end{minipage}
   \caption{{\it(a)} Ratio $\Phi_i^j$ between the unresolved and
   total energies for the ($i$) velocity component along the
   direction ($j$), as a function of $y$.
   \circle, $\Phi_u^x$;
   \trian, $\Phi_w^z$.
   {\it(b)} Fraction $\Theta_i^j$ of the energy of the
   derivative of the ($i$) velocity component with
   respect to the direction ($j$), which is aliased along that same
   direction, plotted as a function of $y$.
   \circle, $\Theta_v^x$;
   \trian, $\Theta_u^z$.
   In all cases, the open symbols refer to the present $Re_\tau= 550$
   simulation, and the closed ones to the one by \cite{MosKimMan99} at
   $Re_\tau= 590$.
}
   \label{fig:alibox}
\end{figure}

The longest scales in the numerical channels occur in the streamwise
velocity $u$ at $y \approx 0.5 h$.  Fig.  \ref{fig:sp1d}{\it (a)} displays
the premultiplied one-dimensional spectra $k_x E_{uu}^{1D}(k_x)$ at that
level.  It is clear that the most energetic structures have lengths of
$2-5h$, which are represented by the Fourier modes $5-12$ in our
simulation at $Re_\tau=550$, and $7-18$ in the one at $Re_\tau=180$.  The
dynamics of the first few Fourier modes are affected by the periodicity of
the box, essentially because their resolution in wavelength space is too
coarse to provide a healthy interaction amongst the different length
scales.  The even-odd structure of the long-wave end of the $1-D$ spectrum
at $Re_\tau=550$ in Fig.  \ref{fig:sp1d}{\it (a)} is probably due to this
effect.  It also appears at other wall distances, and has been observed in
numerical channels performed with completely different numerics (Guglielmo
Scovazzi, private communication).

The widest scales appear at the center of the channel in the spanwise
velocity $w$, whose transverse one-dimensional spectra $k_z
E_{ww}^{1D}(k_z)$ have been represented in Fig.  \ref{fig:sp1d}{\it (b)}.
In this case the peaks of the spectra are sharper than those in Fig.
\ref{fig:sp1d}({\it a}).  Thus, although the most energetic structures are
again associated to low Fourier modes ($6-13$ at $Re_\tau=550$ and $5-9$ at
$Re_\tau=180$), there is relatively less energy in the poorly-represented
modes than in the streamwise direction.

A more quantitative check of the adequacy of the numerical box is to
calculate the fractions $\Phi_{u}^x$ and $\Phi_{w}^z$ respectively of the
streamwise energy $\bra u'^2\ket$ contained in the Fourier modes $k_x=0,\,
k_z \ne 0$, and of the spanwise energy $\bra w'^2\ket$ contained in the
Fourier modes $k_x \ne 0,\, k_z = 0$.  These ratios give an idea of how
much turbulent energy is contained in fluctuations which are longer or
wider than the numerical box, and which are treated numerically as if they
were {\it uniform} in $x$ or $z$.  In Fig.  \ref{fig:alibox}({\it a}) we
represent $\Phi_u^x$ and $\Phi_w^z$ from our DNS at $Re_\tau=550$ and from
\cite{MosKimMan99}.  Note that in the latter, with $L_x \times L_z=2\pi h
\times \pi h$, roughly 20\% of the energy of $u$ is contained in structures
which are longer than the numerical box, and that the behavior of $w$ in
$z$ is even worse in the outer region, where 30\% of its energy is
unresolved.  From these data we conclude that the box of that simulation is
too small to represent the largest flow structures, and that even the
present one is in some ways marginal.  It should however be noted that
numerical experiments at $Re_\tau=180$ in a shorter box with $L_x=8\pi$,
instead of $12\pi$, showed very little degradation in the resolved part of
the longitudinal spectra.  On the other hand, low resolution experiments in
a box of length $L_x=6\pi$ at the higher Reynolds number showed signs of
contamination of the spectral peak by the numerical effects mentioned above
for the long spectral modes.

The grid resolution, given in table \ref{tab:otherguys}, is intermediate
between those used by \cite{MosKimMan99} for their cases $Re_\tau=180$ and
$Re_\tau=590$, and is slightly marginal for the smallest scales, specially
in the $x$ direction.  The result is a spurious accumulation of enstrophy
in the short-wavelength tails of the spectra of the velocity derivatives,
where they are improperly represented.  The most underresolved
derivatives are $\partial_x v$ in $x$ and $\partial_z u$ in $z$.  Fig.
\ref{fig:alibox}{\it (b)} displays the fractions $\Theta_v^x$ and
$\Theta_u^z$ of the enstrophy contained in these underresolved tails in a
way similar to Fig.  \ref{fig:alibox}{\it (a)}.  The underresolved
enstrophy is in this case defined as the integral of the `hook' in the
premultiplied spectrum of the derivative in question, from the highest
wavenumber to the location of its first minimum.  There is more or less
five times more underresolved enstrophy in our numerical channel that in
the one from \cite{MosKimMan99}.  The comparison of Figs.
\ref{fig:alibox}{\it (a)} and \ref{fig:alibox}{\it (b)} shows that the
improperly resolved enstrophy at the short-wave ends of the spectra is of
the same order as that of the improperly resolved energy in their long-wave
ends.

To achieve stationary statistics for structures of wavelength $\lambda$ the
simulation has to be run for several turnover times $\lambda / u_\tau$,
which becomes fairly expensive in these long boxes.  Our experience with
test cases at $Re_\tau=180$ indicates that to have some confidence in the
statistics of the largest scales the simulation should be run for roughly
10 wash-out times $L_x/U_b$, where $U_b$ is the bulk mean velocity of the
flow.  The statistics presented here for the $Re_\tau=550$ case have been
collected during $10$ wash-out times, after discarding initial transients.

\section{Results}

\begin{figure}
  \begin{minipage}[t]{0.48\textwidth}

\includegraphics[height=55mm,width=\textwidth]{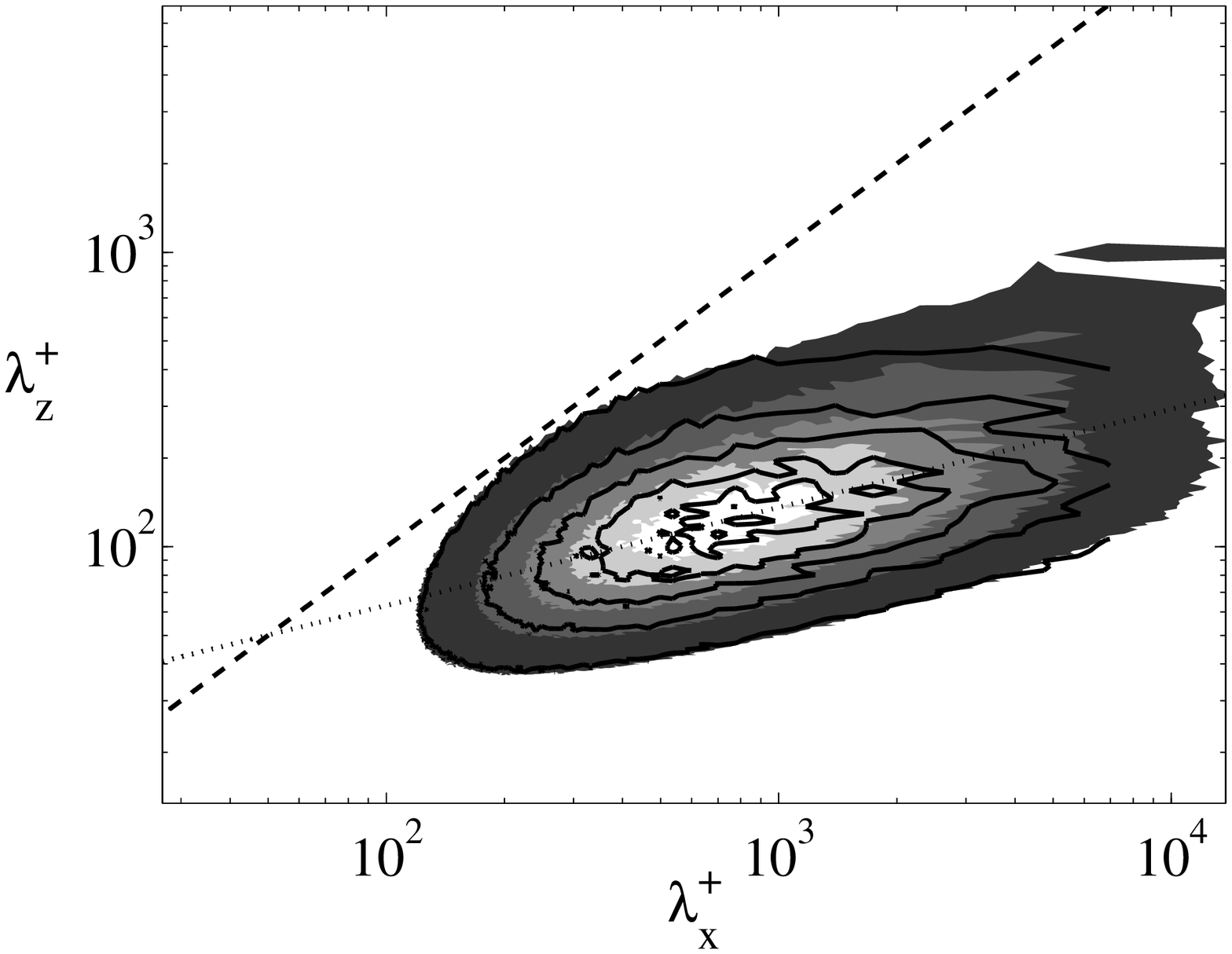}
     \mylab{.2\textwidth}{.8\textwidth}{\it (a)}%
  \end{minipage}
  \hfill
   \begin{minipage}[t]{0.48\textwidth}

\includegraphics[height=55mm,width=\textwidth]{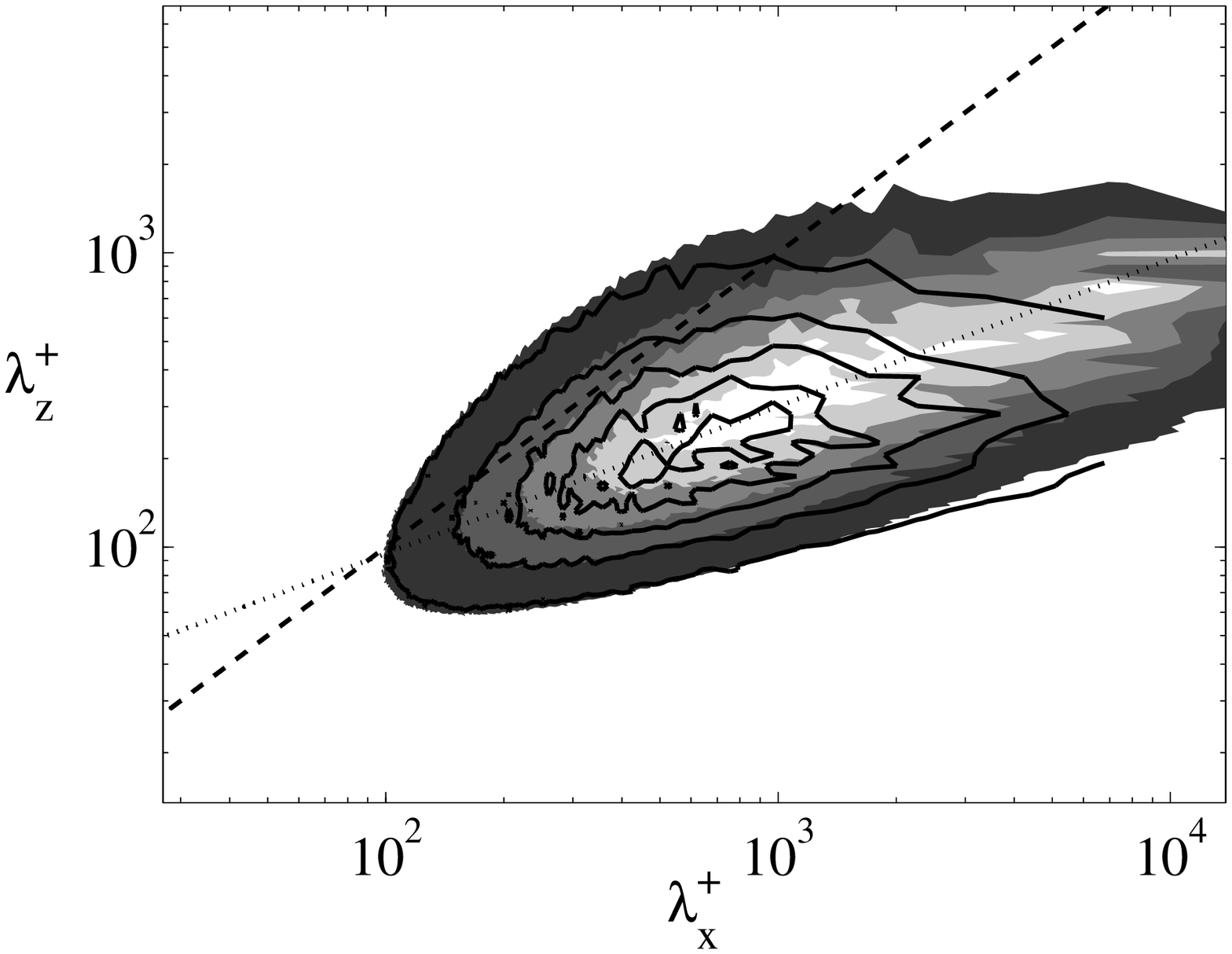}
      \mylab{.2\textwidth}{.8\textwidth}{\it (b)}%
   \end{minipage}
   \begin{minipage}[t]{0.48\textwidth}

\includegraphics[height=55mm,width=\textwidth]{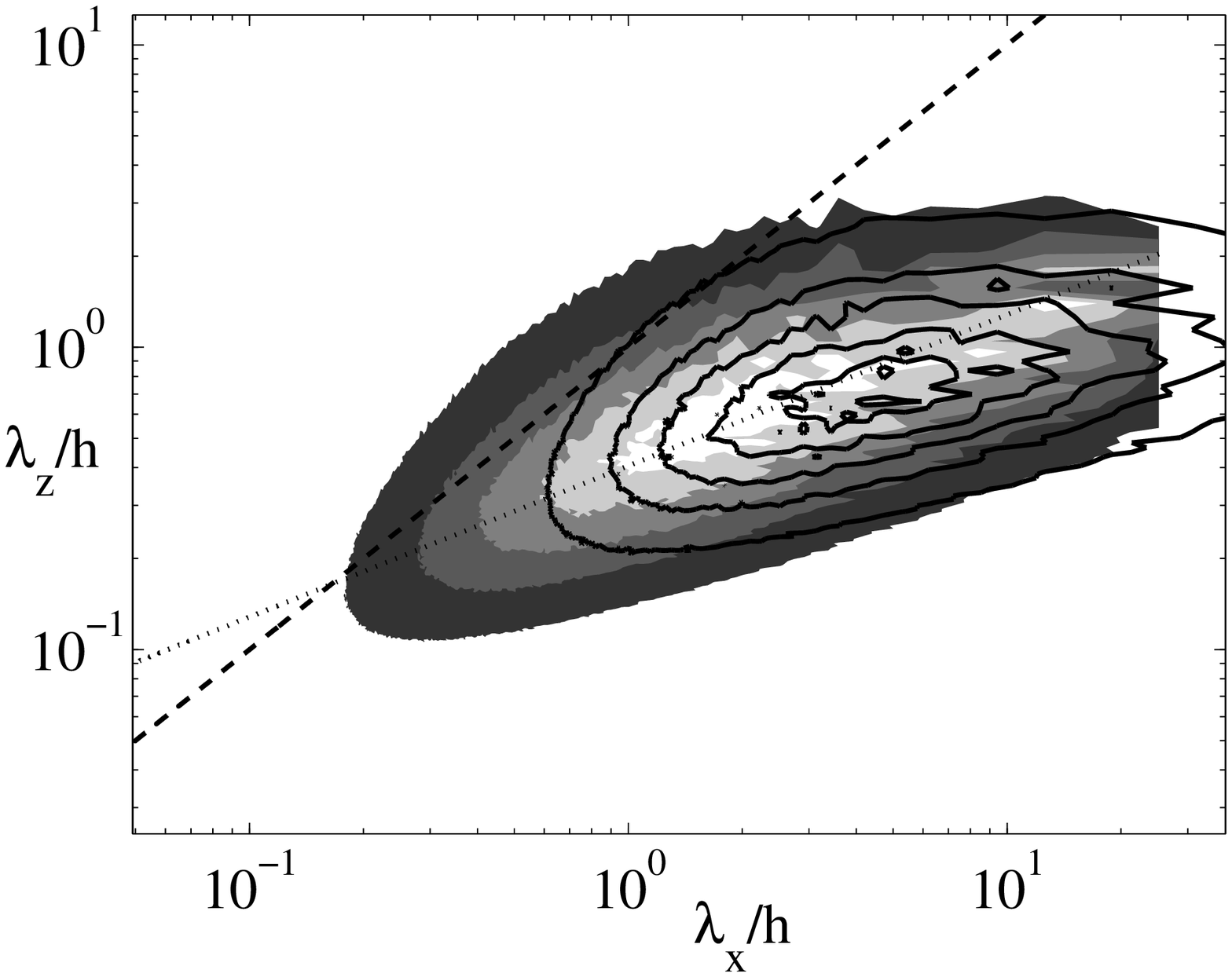}
      \mylab{.2\textwidth}{.8\textwidth}{\it (c)}%
   \end{minipage}
   \hfill
   \begin{minipage}[t]{0.48\textwidth}

\includegraphics[height=55mm,width=\textwidth]{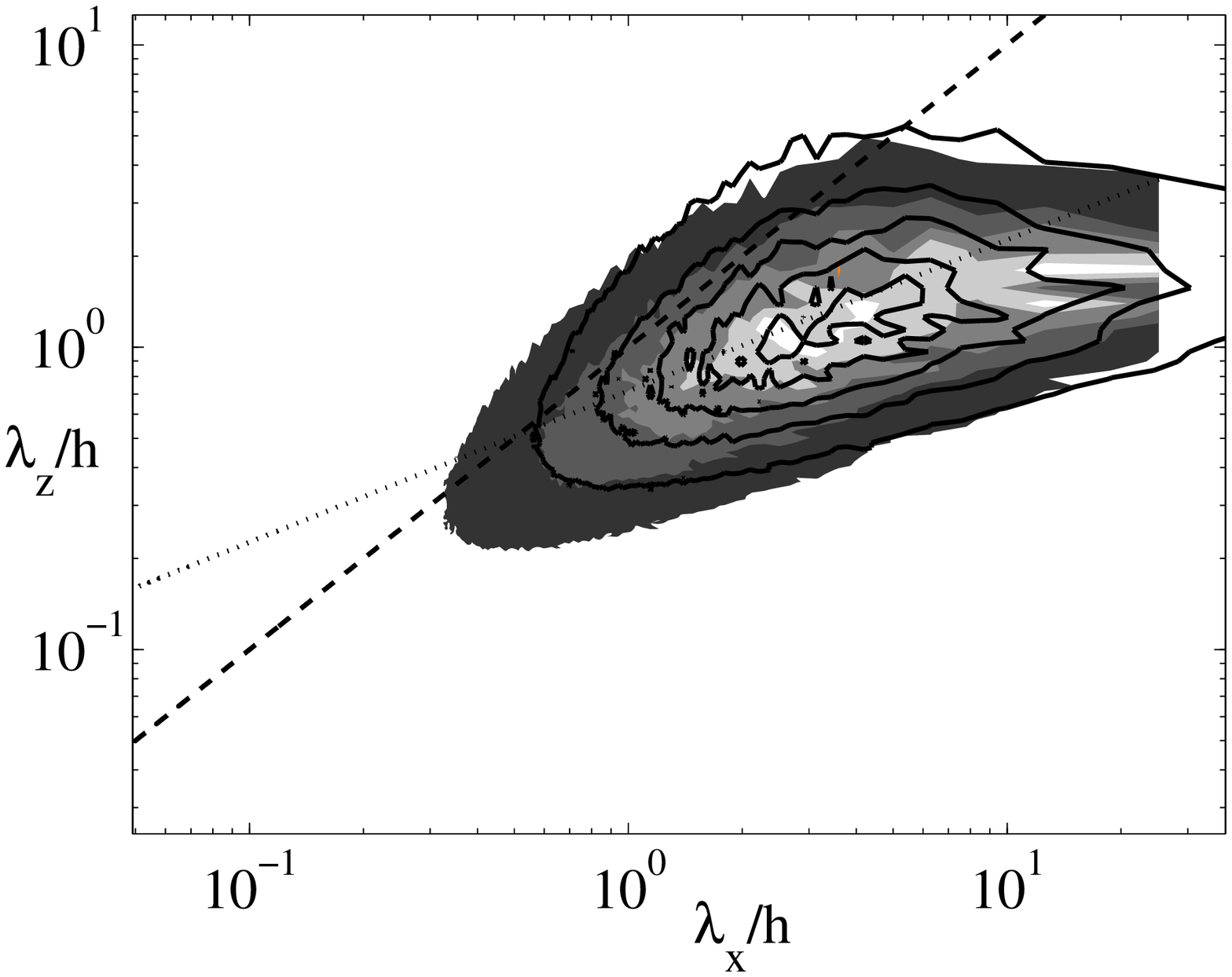}
       \mylab{.2\textwidth}{.8\textwidth}{\it (d)}%
   \end{minipage}
   \caption{Premultiplied two-dimensional spectra
   $\phi_{uu}$ of the streamwise velocity, as functions of the
   streamwise and spanwise wavelengths at three representative wall
   distances.
   {\it (a)} Wall units, $y+=15$;
   {\it (b)} Wall units, $y+=90$ ($y=0.5h$ at $Re_\tau=180$);
   {\it (c)} Outer units, $y=0.2h$ ($y^+=90$ at $Re_\tau=550$);
   {\it (d)} Outer units, $y=0.5h$.
Shaded contours, $Re_\tau=550$; line contours, $Re_\tau=180$.  In all the
cases there are five linearly increasing contours.  \dashed, locus of
two-dimensional isotropic structures $\lambda_z = \lambda_x$; the dotted
line in {\it (a)} is $\lambda_x^+ \sim (\lambda_z^+)^3$, passing through
$\lambda_x^+ = \lambda_z^+ = 50$; those in {\it (b), (c)} and {\it (d)} are
$\lambda_x y = \lambda_z^2$, and the point where this line crosses the
dashed one corresponds to three-dimensionally isotropic structures.
    }
   \label{fig:uu}
\end{figure}

\begin{figure}
  \begin{minipage}[t]{0.48\textwidth}

\includegraphics[height=55mm,width=\textwidth]{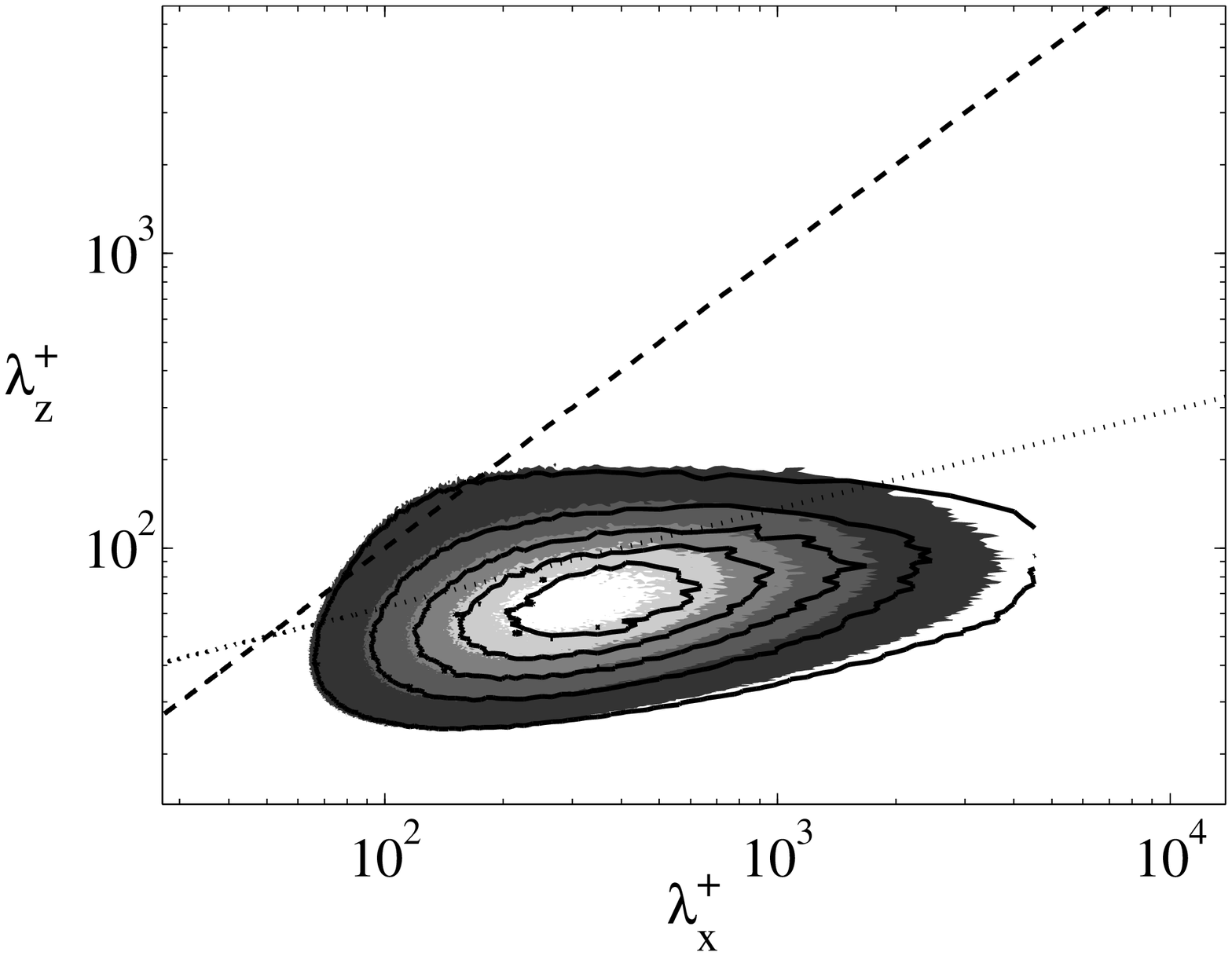}
      \mylab{.2\textwidth}{.8\textwidth}{\it (a)}%
   \end{minipage}
   \hfill
   \begin{minipage}[t]{0.48\textwidth}

\includegraphics[height=55mm,width=\textwidth]{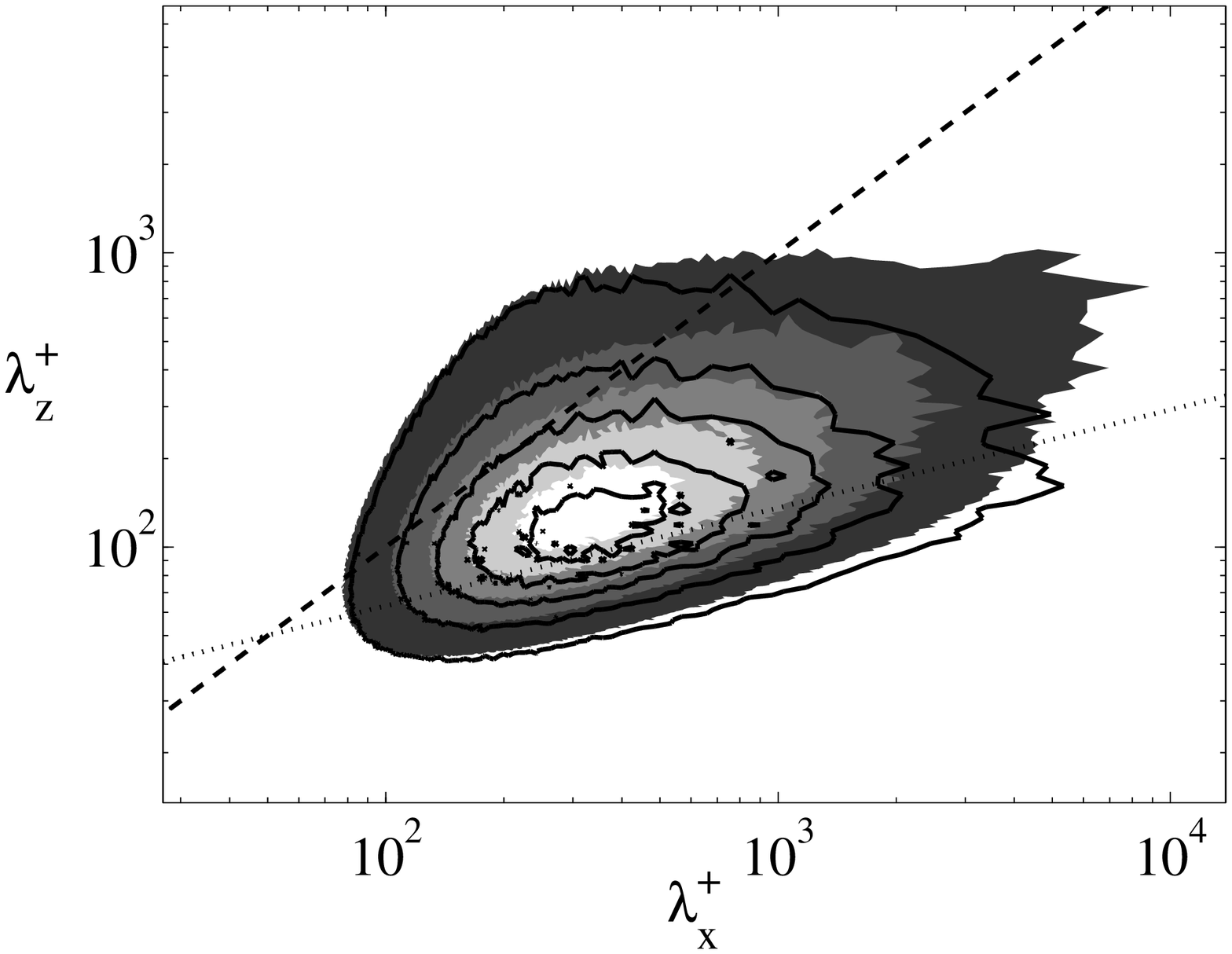}
       \mylab{.2\textwidth}{.8\textwidth}{\it (b)}%
   \end{minipage}
   \begin{minipage}[t]{0.48\textwidth}

\includegraphics[height=55mm,width=\textwidth]{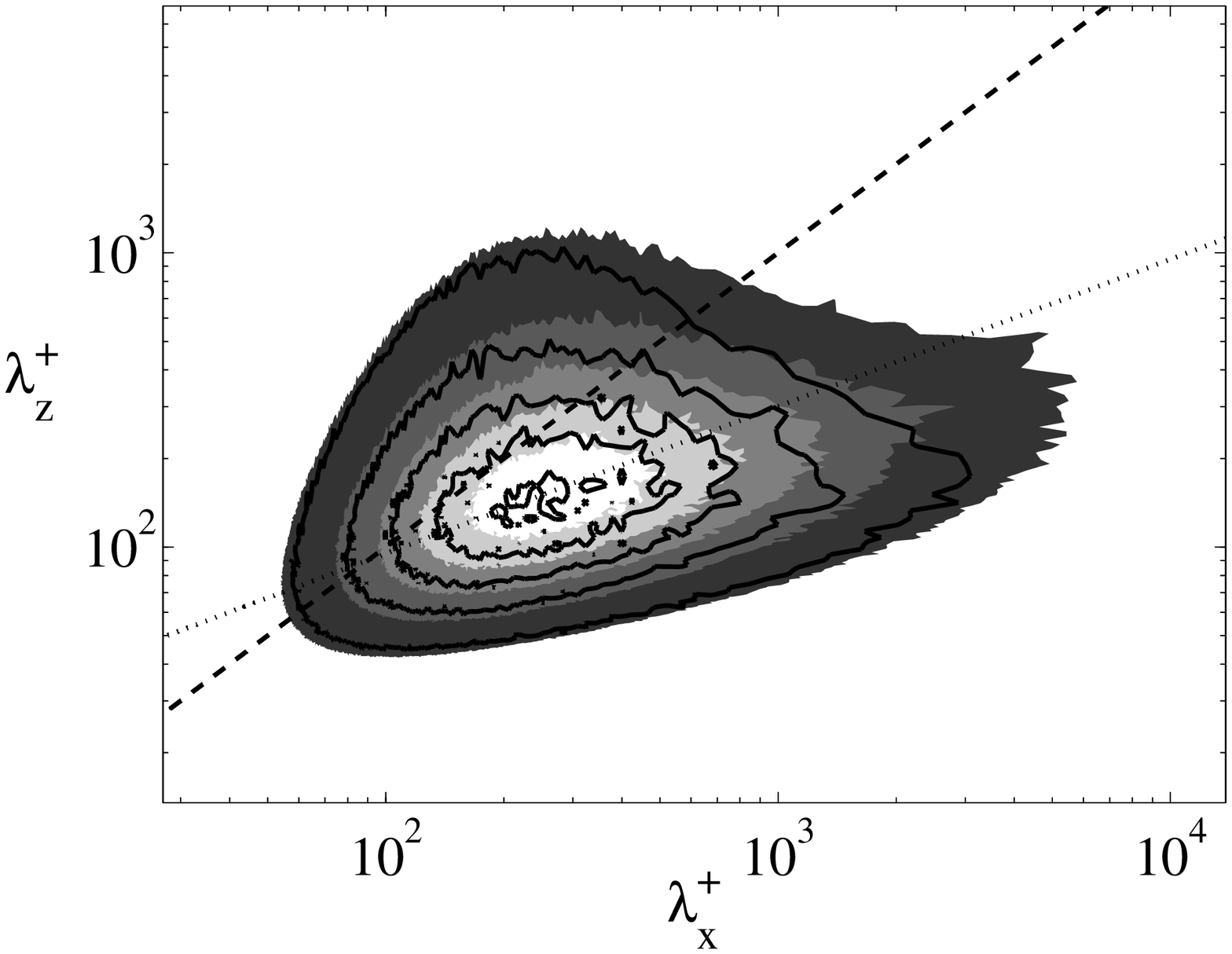}
       \mylab{.2\textwidth}{.8\textwidth}{\it (c)}%
   \end{minipage}
   \hfill
   \begin{minipage}[t]{0.48\textwidth}

\includegraphics[height=55mm,width=\textwidth]{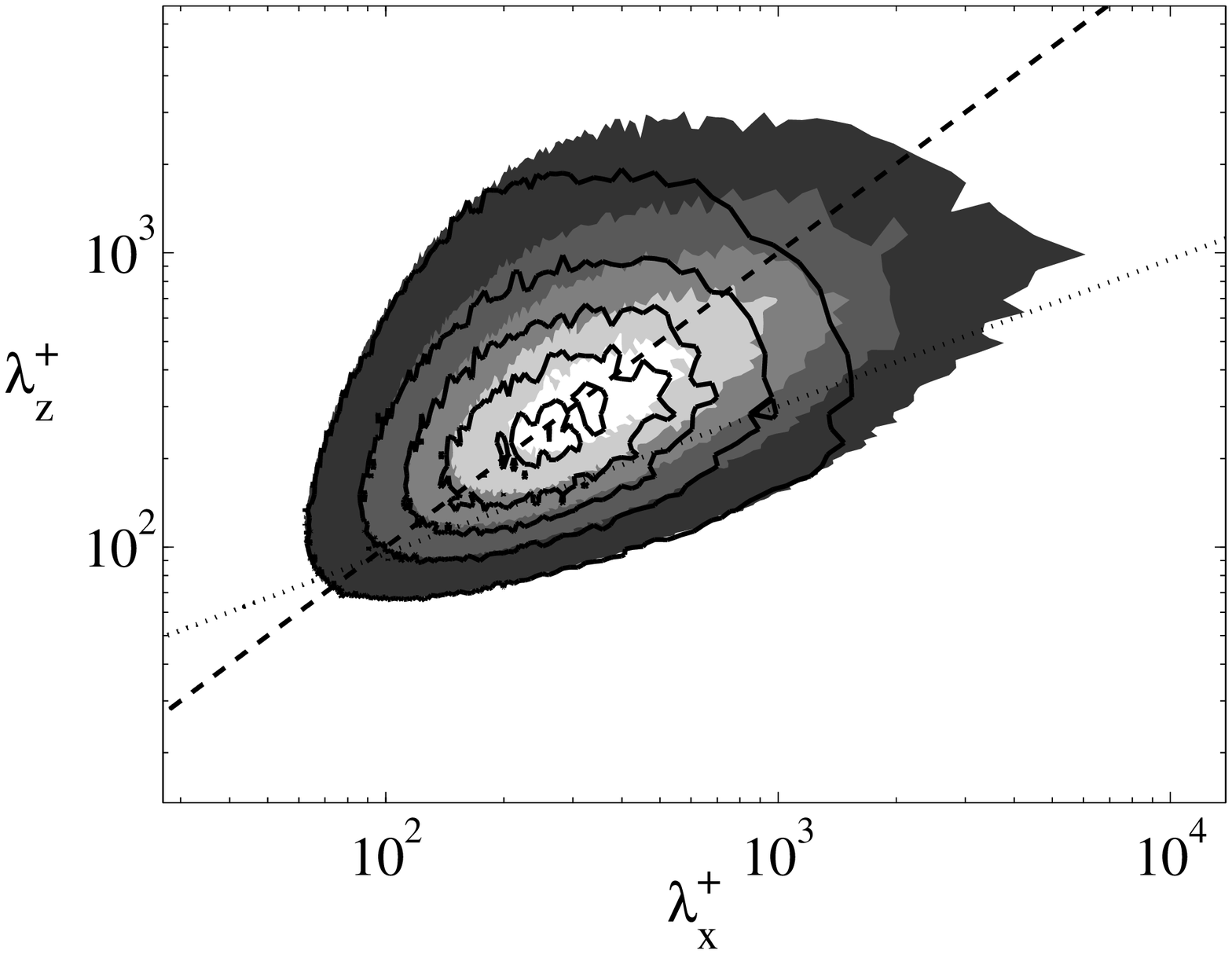}
       \mylab{.2\textwidth}{.8\textwidth}{\it (d)}%
   \end{minipage}
   \begin{minipage}[t]{0.48\textwidth}

\includegraphics[height=55mm,width=\textwidth]{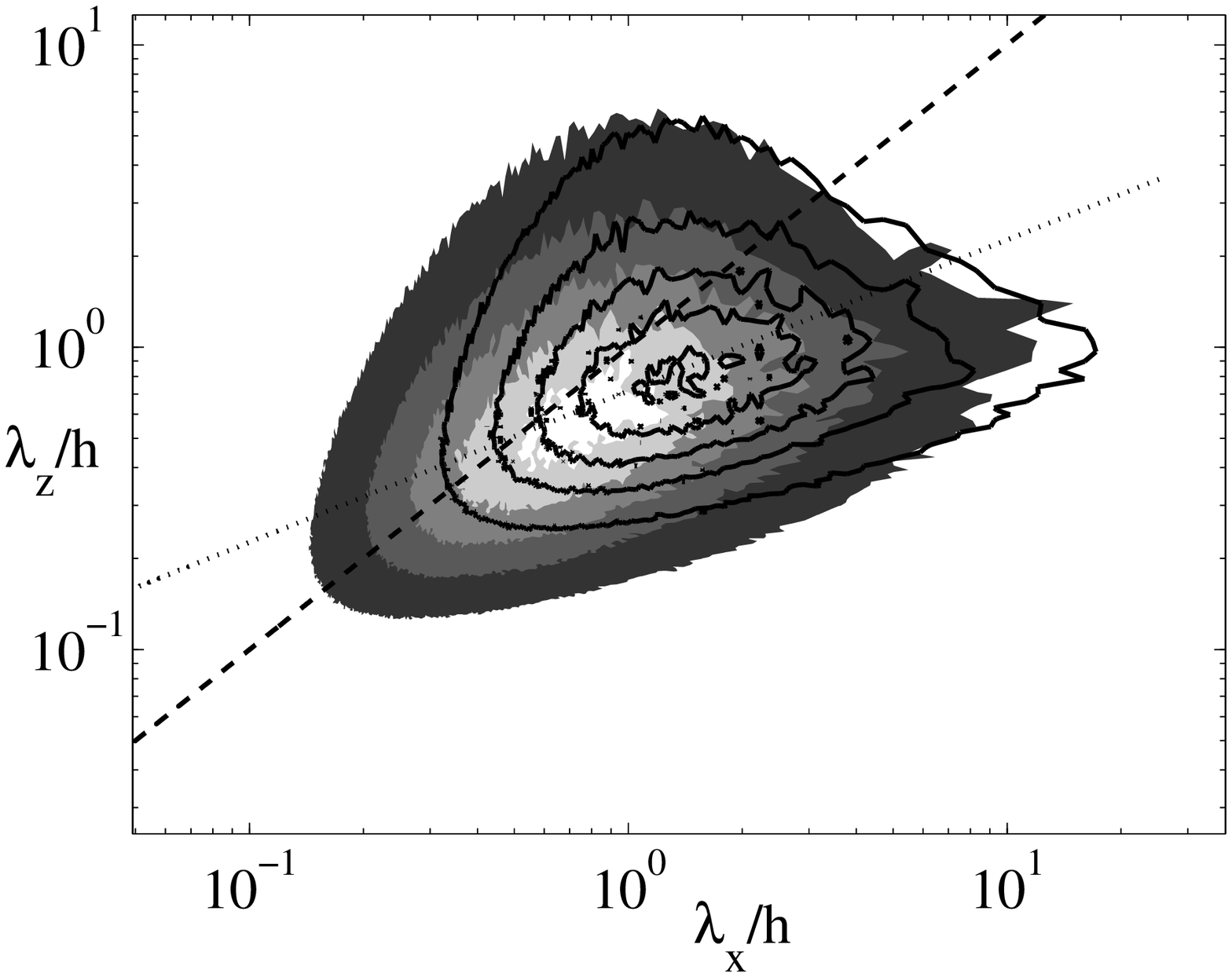}
       \mylab{.2\textwidth}{.8\textwidth}{\it (e)}%
   \end{minipage}
   \hfill
   \begin{minipage}[t]{0.48\textwidth}

\includegraphics[height=55mm,width=\textwidth]{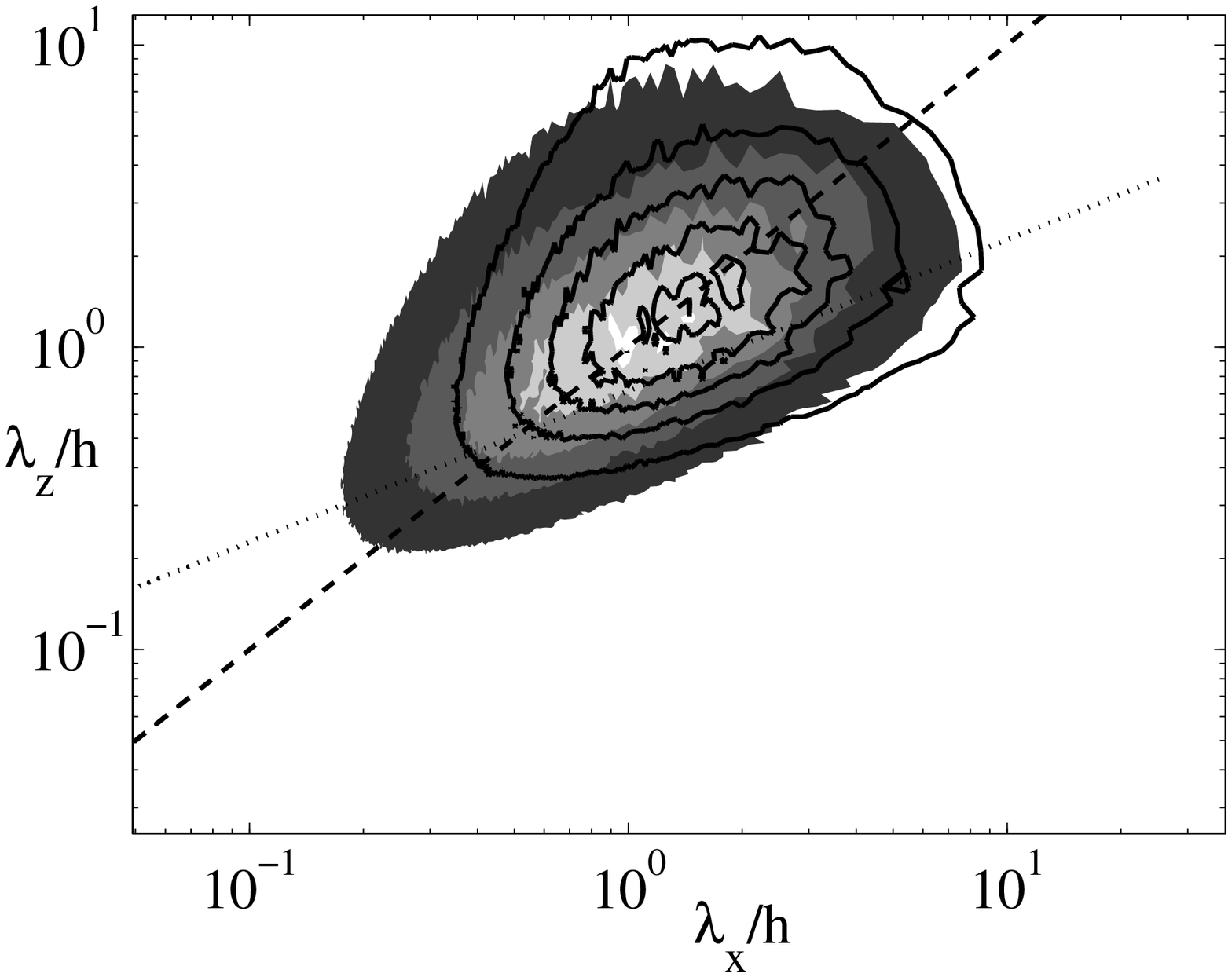}
       \mylab{.2\textwidth}{.8\textwidth}{\it (f)}%
   \end{minipage}
   \caption{Premultiplied two-dimensional spectra as functions of the
   streamwise and spanwise wavelengths at three representative wall
   distances.
   {\it (a), (c) (e)}, $\phi_{vv}$;
   {\it (b), (d) (f)}, $\phi_{ww}$.
   {\it (a), (b)} Wall units, $y+=15$;
   {\it (c), (d)} Wall units, $y^+=90$ ($y=0.5h$ at $Re_\tau=180$);
   {\it (e), (f)} Outer units, $y=0.5h$.
Shaded contours, $Re_\tau=550$; line contours, $Re_\tau=180$.  In all the
cases there are five linearly increasing contours.  \dashed, locus of
two-dimensional isotropic structures $\lambda_z = \lambda_x$; the dotted
lines in {\it (a)} and {\it (b)}, are $\lambda_x^+ \sim (\lambda_z^+)^3$
passing through $\lambda_x^+ = \lambda_z^+ = 50$; those in {\it (c), (d),
(e)} and {\it (f)} are $\lambda_x y = \lambda_z^2$ and the point where both
lines cross in those figures corresponds to three-dimensionally isotropic
structures.
 }
   \label{fig:vv_ww}
\end{figure}

\subsection{Two-dimensional velocity spectra in the near-wall region}

Figs.  \ref{fig:uu} and \ref{fig:vv_ww} display linearly spaced isocontours
of the premultiplied two-dimensional energy spectra $\phi_{ij} = k_x k_z
E_{ij}(\lambda_x,\lambda_z,y)$ as functions of the wavelength vector
$(\lambda_x,\lambda_z) = (2\pi/k_x,2\pi/k_z)$.  Note that
\begin{equation}
\bra u'_iu'_j\ket  \;=\; \int_0^\infty \int_0^\infty
\phi_{ij}(\lambda_x,\lambda_z,y)
\dd (\log \lambda_x) \dd (\log \lambda_z),
\label{eq:intpow}
\end{equation}
so that these figures express how much energy is contained in structures of
length $\lambda_x$ and width $\lambda_z$.  The shaded contours come from
the simulation at $Re_\tau=550$, while the line contours are from the one
at $Re_\tau=180$.  The four wall distances in Fig.  \ref{fig:uu} are
$y^+=15,\;y^+=90,\;y=0.2h,\;y=0.5h$, corresponding respectively to the
bottom and the top of the buffer layer, and the bottom and the core of the
outer region.  In Fig.  \ref{fig:vv_ww} the three wall distances are
$y^+=15,\;y^+=90$ and $y=0.5h$.

In the wall region the spectrum of the streamwise velocity (Fig.
\ref{fig:uu}{\it a}) peaks around $\lambda_x^+ \approx 700,\, \lambda_z^+
\approx 100$, which is the size of the buffer layer streaks.  The spectra
of the two other velocity components peak around $\lambda_x^+ \approx
250,\, \lambda_z^+ \approx 50-100$ (see Figs.  \ref{fig:vv_ww}{\it a} and
\ref{fig:vv_ww}{\it b}), corresponding approximately to the dimensions of
an individual system of counterrotating quasi-streamwise vortices (\KMM87).

There is still not general agreement about the scaling in the near-wall
region.  Contrary to the classical idea that inner scaling should work
close enough to the wall, several experimentalists have found evidence
suggesting that this is not so, and in particular that the streamwise
normal stress $\bra u'^2\ket $ increases with the Reynolds number
throughout the wall layer, when expressed in wall units at a fixed $y^+$.
Some of these researchers (\GE00; \PL90) have argued that the Reynolds
number dependence is due to the contribution of Townsend's (1976)
`inactive' motions.  They note that this contribution scales in outer
units, and are motivated by this observation to introduce `mixed' scaling
in which $\bra u'^2\ket$ is proportional to the geometric mean of the friction and
outer velocities.  \cite{Hit97} presents a similar argument, but favors an
interpretation in which the inner and outer contributions are scaled
independently, with no simple overall law.  \cite{JimFloMan01}, using data
from \cite{Hit97} and from the present numerical simulations show that the
short-wavelength end of the one-dimensional streamwise $u$ spectrum at
$y^+=20$ scales well in inner units, while its long-wavelength end scales
in outer units.  It was already noted by \cite{Tow76} that the no-slip
impermeability condition at the wall does not limit the size of the $u$ and
$w$ velocity structures, while the effect of the no-slip condition is
limited in height to a few wall units.  We can therefore expect the large
scales, even if they originate far from the wall, to penetrate deep into
the near-wall layer, causing $\bra u'^2\ket $ to have both local and global
contributions.

In fact, the only region of the two-dimensional $u$-spectra in Fig.
\ref{fig:uu}({\it a}) that does not collapse well in wall units is the
upper-right corner, which corresponds to the large structures which
dominate the outer-layer spectra in Figs.  \ref{fig:uu}{\it (c)} and
\ref{fig:uu}{\it (d)}.  As we move deeper into the buffer layer the energy
contained in the large scales increases (Fig.  \ref{fig:uu}{\it b}), in
agreement with the common observation that the collapse of $\bra u'^2\ket $
in inner scaling worsens as the wall distance increases.  Recent
experimental spectra by \cite{MetKle01} at $y^+=15$ in the atmospheric
boundary layer show that a substantial fraction of the streamwise turbulent
energy is contained in very large structures at those extremely high
Reynolds numbers ($Re_\theta \sim 10^6$).  \cite{JimFloMan01} have argued
that the effect is actually repressing, with the outer large scales
preventing the wall streaks from becoming `infinitely' long.

It is worth pointing out that the $u$ spectrum in this region lies
approximately along the power law
\begin{equation}
\lambda_x^+ \;\sim\; \left({\lambda_z^+}\right)^3,
\label{eq:powlawwall}
\end{equation}
implying that, while the structures of the streamwise velocity become wider
as the become longer, they also become more elongated, since they
progressively separate from the spectral locus of two-dimensional isotropy.
The spectrum of $w$ does not share this property and is more isotropic in
the $(x,z)$ plane.  The spectrum of $v$ is very anisotropic in the
near-wall region, but as we move away from the wall it develops a second
isotropic component (Fig.  \ref{fig:vv_ww}{\it c,e}) whose relative
strength increases with the wall distance, and which becomes dominant close
to the center of the channel.

\subsection{The very large anisotropic scales in the outer layer}

Above $y^+ \approx 60$, the spectrum of the streamwise velocity becomes
quite different from the spectra of the wall-normal and spanwise
velocities, as we can see in Figs.  \ref{fig:uu} and \ref{fig:vv_ww}.  The
spectrum of $u$ is anisotropic and has two components.  The first one is
associated to small scales, and collapses fairly well for our two Reynolds
numbers when plotted as a function of ($\lambda_x^+,\;\lambda_z^+$) at a
constant $y^+$ (Figs.  \ref{fig:uu}{\it a} and \ref{fig:uu}{\it b}).  The
second one is related to large anisotropic structures and collapses well
when plotted at a constant $y/h$ as a function of ($\lambda_x
/h,\,\lambda_z /h$) (Figs.  \ref{fig:uu}{\it c} and \ref{fig:uu}{\it d}).
Below $y^+ \approx 60$ the small-scale component is the most important one,
and the peak of the spectrum collapses in wall units as in Fig.
\ref{fig:uu}{\it (a)}.  Far from the wall ($y \gtrsim 0.3h$), it is the
large-scale component which dominates, and the peak of the spectrum
collapses in outer units as in Fig.  \ref{fig:uu}{\it (d)}.  This
description suggests that, at least in turbulent channels at moderate
Reynolds numbers, there exists a family of $u$ structures in the inner
region which scales in wall units and another one in the outer region which
scales in outer units.

The present results resemble those of the high-Reynolds number experiments
of \cite{Hit97} and \cite{KimAdr99}, although there is a significant
difference.  In the numerical channels the two spectral peaks corresponding
to the VLAS and the wall streaks never coexist.  Instead, we observe an
intermediate region of wall distances ($y^+ \gtrsim 60,\, y \lesssim 0.3h$)
where the peaks of the $u$ spectra do not collapse in inner or in outer
units (Figs.  \ref{fig:uu}{\it c} and \ref{fig:uu}{\it d}).  One reason for
this discrepancy may be that the viscous and the large-scale components of
the $u$ spectra have comparable intensities in our moderate-Reynolds number
simulations.  \cite{huntm} have noted that the energy contained in the
outer structures increases with the Reynolds number and will eventually
become much larger than that in the inner ones, so that the large scales
would eventually become dominant even very near the wall.  The failure to
observe an overlap in our simulations could be related to that effect;
although we clearly observe the two spectral components in the
two-dimensional spectra, the outer one is never strong enough in the inner
region to appear as a peak in the one-dimensional spectra.  On the other
hand, new measurements at extremely high Reynolds numbers by
\cite{Moretal01} in pipes do not show any region in the flow with double
spectral peaks for the one-dimensional $u$-spectra, and the question should
therefore be consider as still open.

The spectra of $v$ and $w$ are also in this region more isotropic than
those of $u$.  The spectrum for $w$ is closer to two-dimensional isotropy
than that of $v$, but it is flatter, in the sense that it is both wider and
longer than $v$ for a given height.  It is difficult from the present
results to obtain clear spectral scaling laws for the transverse velocity
components.  In the lower part of the outer layer the small scales of the
spectra collapse well in inner units (Figs.  \ref{fig:vv_ww}{\it c} and
\ref{fig:vv_ww}{\it d}), similarly to what happens with the streamwise
velocity, but in this case the large scales do not collapse in outer units.
Further away from the wall we have been unable to find any scaling that
collapses the spectra at $Re_\tau=180$ with those at $Re_\tau=550$.  In
Figs.  \ref{fig:vv_ww}{\it(e)} and \ref{fig:vv_ww}{\it(f)} the spectra are
represented in outer units and only collapse, and even then imperfectly,
for the largest scales.

\begin{figure}
  \begin{center}
\includegraphics[height=0.4\textwidth]{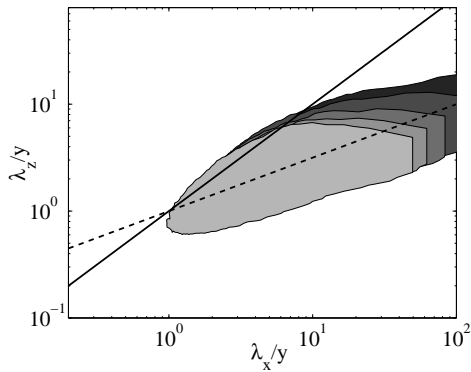}
   \end{center}
   \caption{Superimposed contours of $0.2$ times the maximum of $\phi_{uu}$
   at five wall distances in the outer layer
   (from dark to light $y=0.1h (0.1h) 0.5h$). They are
   represented as functions of the
   streamwise and spanwise wavelengths nondimensionalized with the wall
   distance. \solid, locus of two-dimensional isotropic structures
   $\lambda_z/y = \lambda_x/y$; \dashed, $(\lambda_z/y)^2 = \lambda_x/y$.
   The point where both lines cross
   corresponds to three-dimensionally isotropic structures.}
   \label{fig:lz2lxy}
\end{figure}

The two-dimensional spectrum of the streamwise velocity exhibits a
different behavior in the outer layer than in the wall region (see Eq.
\ref{eq:powlawwall}).  Fig.  \ref{fig:lz2lxy} displays superimposed
contours of $\phi_{uu}$ at five different wall distances in the outer layer
ranging from $y=0.1h$ (light) to $y=0.5h$ (dark).  The spectra are
nondimensionalized with $y$ and collapses well along the dashed line, which
corresponds to the power law
\begin{equation}
\lambda_x  y = {\lambda_z}^2.
\label{eq:lxfromlz}
\end{equation}
A possible explanation for this power law is that the structures in the
streamwise velocity are the decaying wakes of approximately isotropic $v$
and $w$ structures.  Those of diameter $\lambda_z$ decay in times of order
$\lambda_z^2/\nu_T$ under the action of an eddy viscosity $\nu_T$, leaving
`wakes' in the streamwise velocity whose length is
\begin{equation}
\lambda_x \sim U_b\lambda_z^2/\nu_T,
\label{eq:lambda2}
\end{equation}
assuming that they are convected at a velocity of the order of the bulk
velocity.  The choice of a constant advection velocity implies that
necessarily the large structures feel the wall, since velocity itself is
not a Galilean invariant.

\begin{figure}
   \begin{center}
\includegraphics[height=0.4\textwidth]{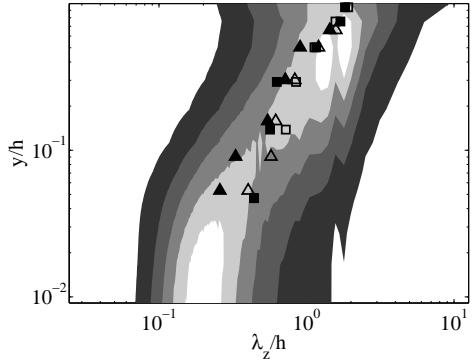}
\caption{Premultiplied 1--D spectrum of streamwise velocity, in outer
units.  The shaded contours are the present $Re_\tau = 550$ simulation; the
symbols are experiments by \cite{NakNez81}.  \solidtrian and \trian,
$Re_\tau=696$; \solidsquar and \squar, $Re_\tau=318$.  Open symbols, $\bra
\lambda\ket_s$; closed symbols, $\bra \lambda\ket_e$.
}
   \label{fig:specy}
   \end{center}
\end{figure}

The relation (\ref{eq:lxfromlz}) not only expresses how $\phi_{uu}$ is
organized in the plane ($\lambda_x, \lambda_z$) at a given wall distance.
Since Eq. (\ref{eq:lambda2}) links the coefficient of the power law in Eq.
(\ref{eq:lxfromlz}) to the magnitude of the eddy viscosity, the fact that
all the spectra in Fig.  \ref{fig:lz2lxy} are aligned along a single line
implies that $\nu_T$ is proportional to $y$, in agreement, and in strong
support, to the similarity arguments about the scaling of the Reynolds
stresses used in the standard derivations of the logarithmic velocity
profile.  It also helps understand why the outer structures become more
isotropic with wall distance (which can be noticed from the displacement of
the contours in figure \ref{fig:lz2lxy}), since the decaying time of the
wakes decreases as the eddy viscosity increases.  It should however be noted
that the eddy viscosity in this flow, as measured from the mean velocity
profile, only increases approximately linearly up to $y\approx 0.2$, and is
constant thereafter, so that most of the spectra in Fig.  \ref{fig:lz2lxy}
are outside the region of linear dependence.  The present model can
therefore only be taken as indicative until detailed calculations are
carried out using the real eddy viscosity distribution.  How it can be
reconciled with the different power law (\ref{eq:powlawwall}) observed near
the wall is briefly discussed in \cite{JimFloMan01}.

The spectra of the three velocity components suggest that the $u$
structures in the outer flow region resemble the buffer layer streaks,
although they differ from them in that they are themselves turbulent, and
in that it is unclear whether they are flanked by quasi-streamwise
vortices.  They seem however to be associated with roughly isotropic
turbulent structures of the transverse velocities whose width increases
with $y$ (Fig.  \ref{fig:vv_ww}), but whose kinematics are unknown.  The
$u$-VLAS also widen with wall distance, specially above the buffer layer.
This is shown in Fig.  \ref{fig:specy}, which displays the transverse
one-dimensional spectrum $k_z E_{uu}^{1D}$ at $Re_\tau=550$.  The spectrum
has been plotted as a function of $\lambda_z$ and $y$, and it has
been non-dimensionalized with the local streamwise energy $\bra u'^2\ket(y)
$. The figure therefore shows how much energy is associated to $u$
structures of a certain width $\lambda_z$ at a given distance to the wall.
The figure also includes the widths of the $u$ structures obtained by
\cite{NakNez81}.  They measured the spanwise organization in a turbulent
open channel with a free surface using the autocorrelation of the
streamwise velocity conditionally averaged with the presence of ejections,
$\bra \lambda\ket_e$, and sweeps, $\bra \lambda\ket_s$.  Their data agree
reasonably well with ours even in the outer region, where the different
geometrical configurations could be expected to affect the nature of the
flow.  The transverse one-dimensional spectra of the transverse velocity
components, not shown here, behave with $y$ very much like those of $u$.

\subsection{The cospectrum}

The cospectrum is particularly important because its integral is the
Reynolds stress $\bra u'v'\ket $, and determines the mean velocity profile
$U$ and the production of turbulent kinetic energy.

\begin{figure}
  \begin{minipage}[t]{.48\textwidth}

\includegraphics[height=55mm,width=\textwidth]{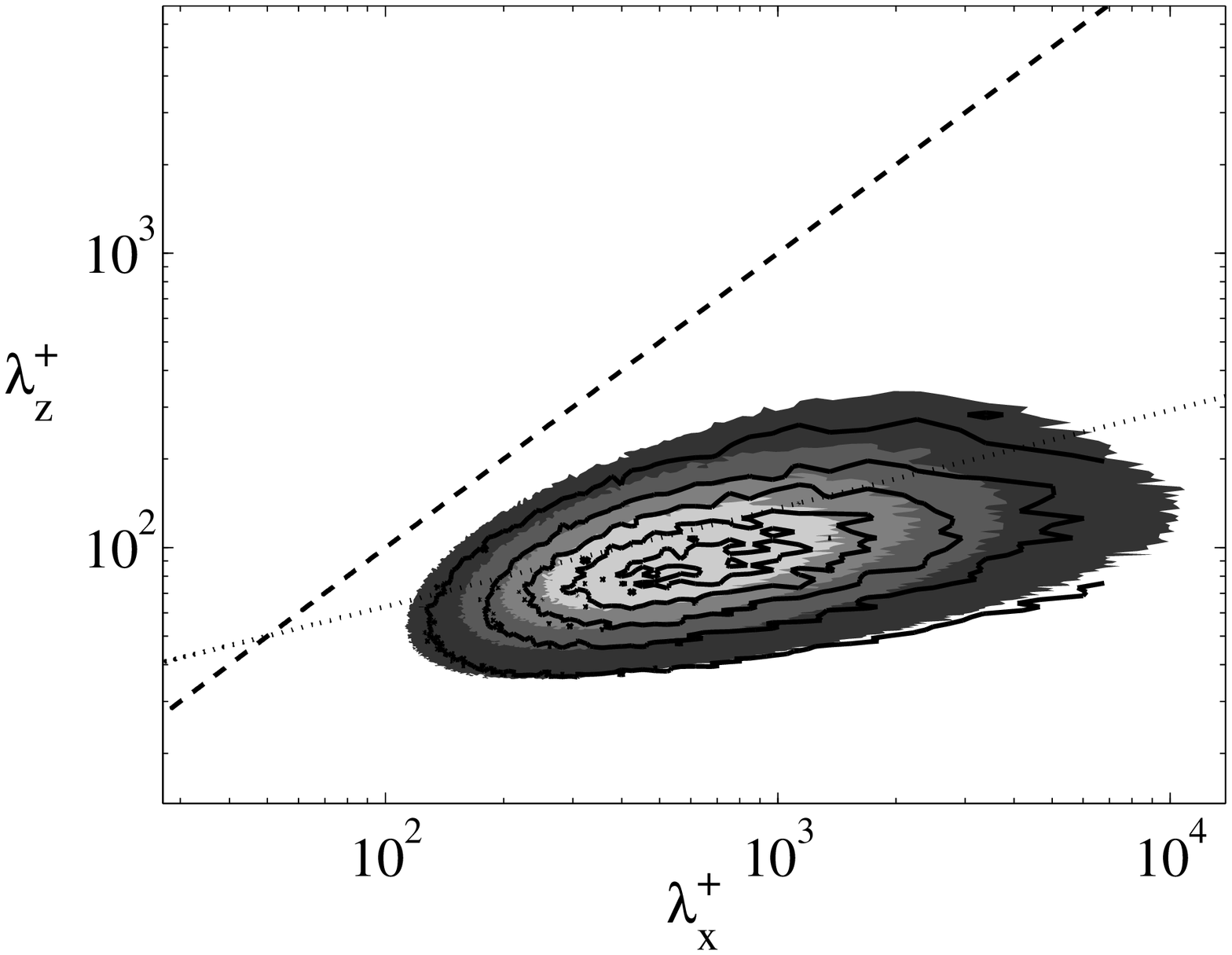}
       \mylab{.2\textwidth}{.8\textwidth}{\it (a)}%
   \end{minipage}
   \hfill
  \begin{minipage}[t]{.48\textwidth}

\includegraphics[height=55mm,width=\textwidth]{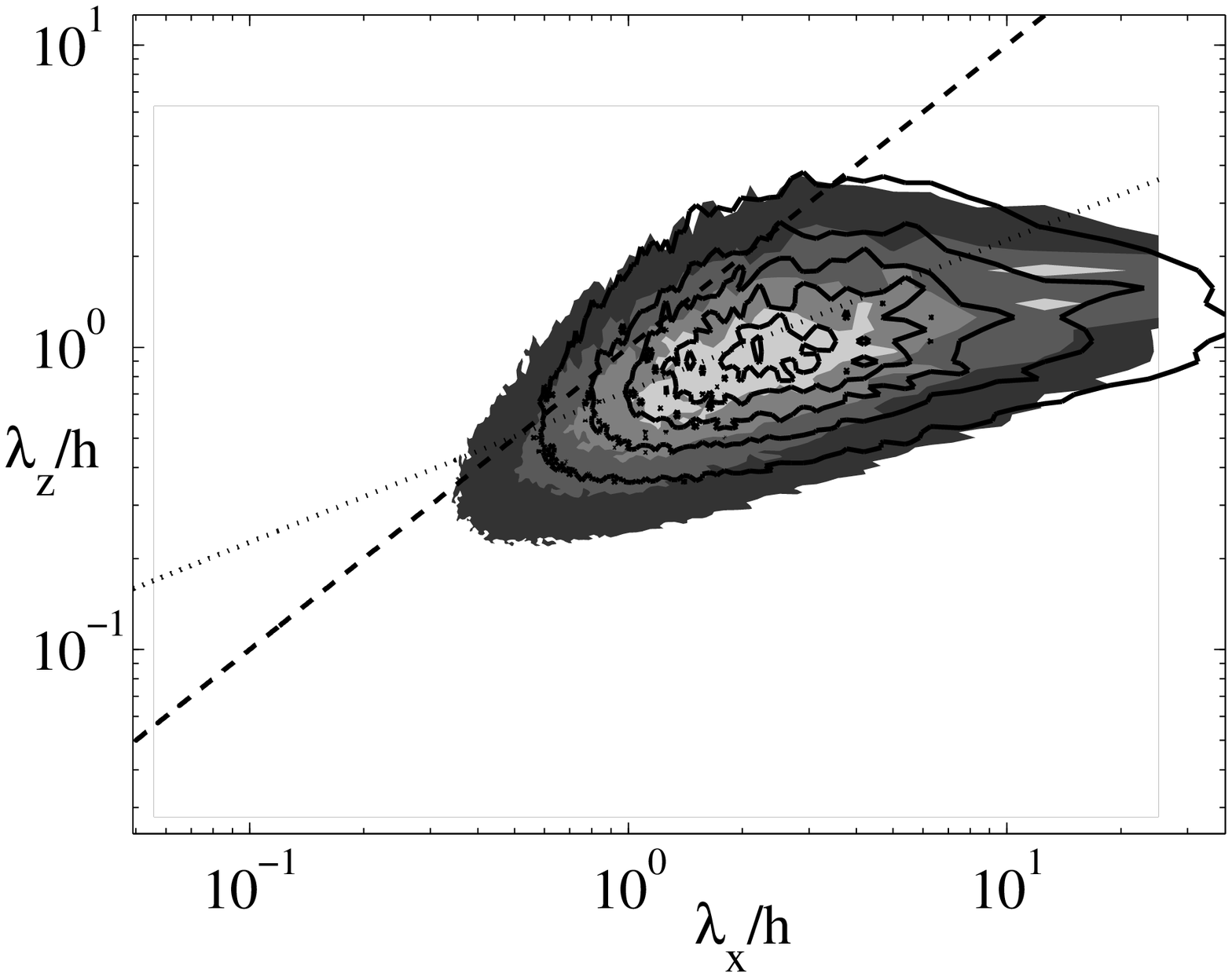}
       \mylab{.2\textwidth}{.8\textwidth}{\it (b)}%
   \end{minipage}
   \caption{Premultiplied two-dimensional cospectra as functions of the
   streamwise and spanwise wavelengths at two representative wall
   distances.
   {\it (a)} Wall units, $y^+=15$;
   {\it (b)} Outer units, $y=0.5h$.
   Shaded contours, $Re_\tau=550$; line contours, $Re_\tau=180$.
   In all the cases there are five linearly increasing contours.
   \dashed, locus of two-dimensional isotropic structures
   $\lambda_z = \lambda_x$; the dotted line in
   {\it  (a)} is $\lambda_x^+ \sim (\lambda_z^+)^3$, passing through
   $\lambda_x^+ = \lambda_z^+ = 50$; that in
   {\it  (b)} is $\lambda_x y = \lambda_z^2$.
   }
   \label{fig:uv}
\end{figure}

Fig.  \ref{fig:uv} shows the premultiplied two-dimensional cospectra in the
near-wall and in the outer regions of the flow in the same fashion as the
two-dimensional premultiplied velocity spectra.  They resemble much more
the premultiplied spectra of $u$ in Figs.  \ref{fig:uu}({\it a}) and
\ref{fig:uu}({\it d}) than those of $v$ in Figs.  \ref{fig:vv_ww}({\it a})
and \ref{fig:vv_ww}({\it e}).  This is true at all wall distances.  The
cospectra collapse in inner scaling in the near-wall region and in outer
scaling in the outer region, in the same way as the $u$-spectra.  But
unlike the latter the cospectra fully collapse in wall units in the
near-wall region, suggesting that the outer large structures do not affect
the Reynolds stresses very close to the wall.  This result agrees with the
experiments by \cite{GraEat00}, who found that $\bra u'v'\ket $ scales in
inner units close to the wall and give support to the law of the wall
$U^+=U^+(y^+)$.  \cite{Jim98} also noted that the one-dimensional
experimental cospectra in the logarithmic layer scaled with $y$ much better
than any of the other available velocity components.

It is of special interest that in the outer region, the VLAS carry a
substantial fraction of the Reynolds stresses even if they are not present
in the spectra of the wall-normal velocity in Fig.  \ref{fig:vv_ww}{\it
(e)}.  \cite{Jim98} observed that the vanishing of the {\it premultiplied}
spectrum of $v$ for long or for wide waves is not sufficient to imply that
the {\it premultiplied} cospectrum also vanishes in that limit, as it had been
previously assumed (\PHC86).  Let
\begin{equation}
E_{uv}\; =\; \sigma_{uv} {\left({E_{uu} E_{vv}}\right)}^{1/2},
\label{eq:strupar}
\end{equation}
where $\sigma_{uv}$ is the structure parameter, which measures the
correlation between $u$ and $v$ and the efficiency of those fluctuations
in transporting momentum.  In the present simulations, $E_{vv}$ is
independent of $\lambda_x$ for the long scales and consequently, the
premultiplied $v$ spectrum
$$
\frac {(2 \pi)^2 E_{vv} } { \lambda_x \lambda_z},
$$
goes to zero as $1/\lambda_x$ when $\lambda_x \gg 1$, while $\phi_{uu}$
decays more slowly.  It then follows from the square root in in Eq.
(\ref{eq:strupar}) that in the limit of very long wavelengths the behavior
of the premultiplied cospectrum depends on the prefactor $\sigma_{uv}$.

\begin{figure}
  \begin{minipage}[t]{.48\textwidth}

\includegraphics[height=55mm,width=\textwidth]{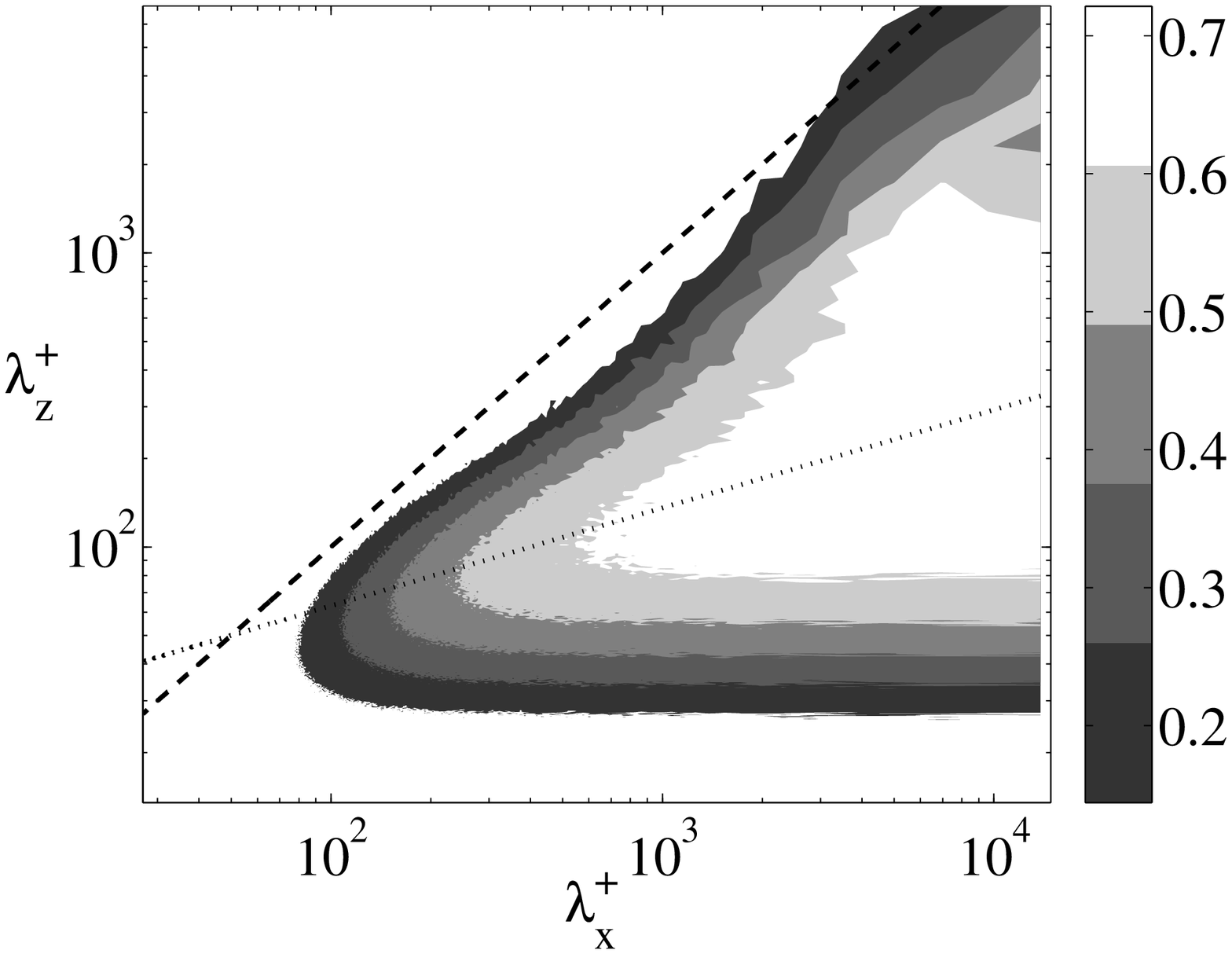}
       \mylab{.2\textwidth}{.8\textwidth}{\it (a)}%
   \end{minipage}
   \hfill
  \begin{minipage}[t]{.48\textwidth}
\includegraphics[height=55mm,width=\textwidth]{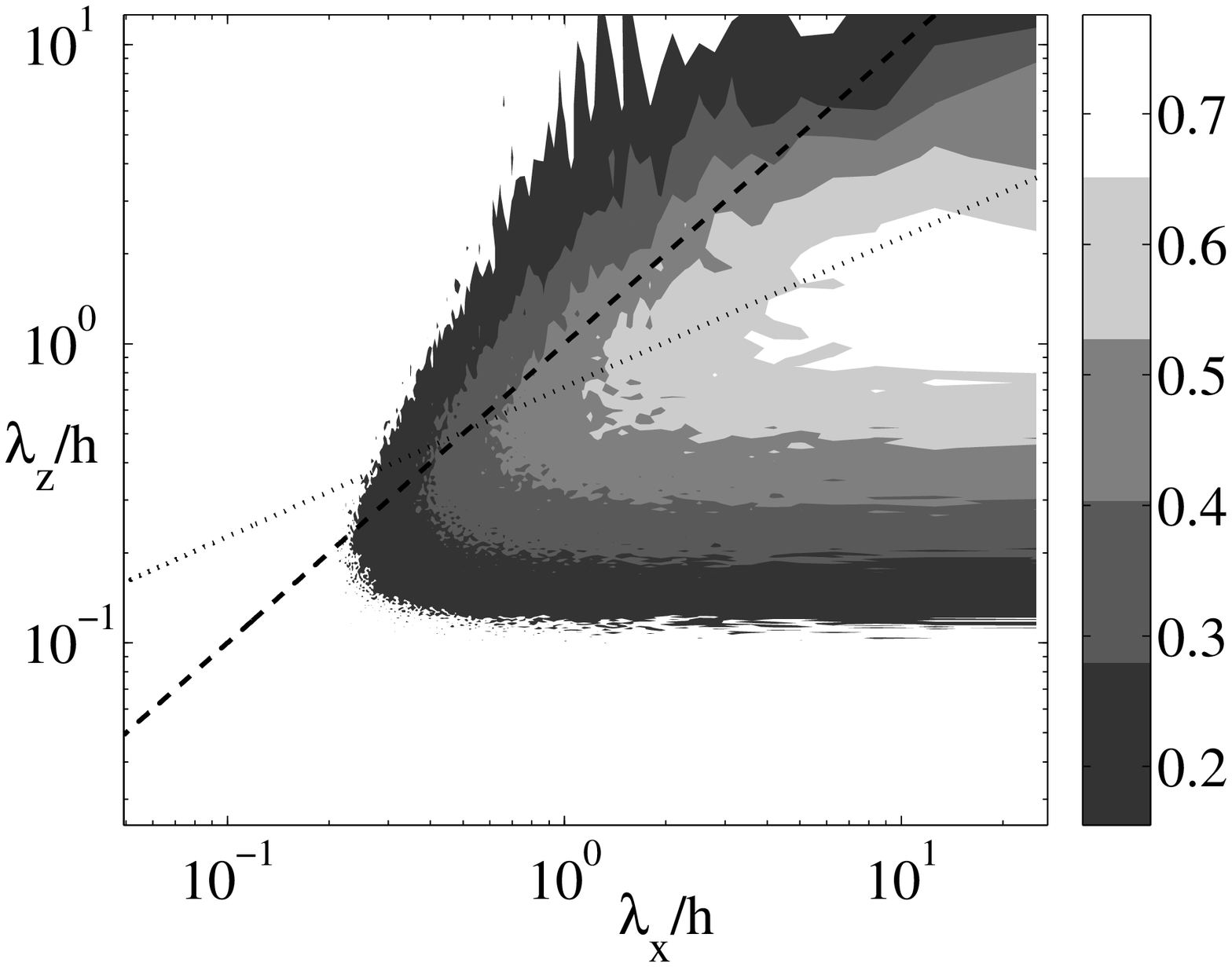}
       \mylab{.2\textwidth}{.8\textwidth}{\it (b)}%
   \end{minipage}
\caption{Structure parameter $\sigma_{uv}$ as a function of the streamwise
and spanwise wavelengths at two representative wall distances.  Present
$Re_\tau=550$.  ({\it a}) wall units, $y^+=15$; ({\it b}) outer units,
$y=0.5h$.  In both cases there are five linearly increasing contours.
\dashed, locus of two-dimensional isotropic structures $\lambda_z =
\lambda_x$; the dotted line in {\it (a)} is $(\lambda_z^+)^3 \sim
\lambda_x^+$ passing through $y^+=50$; that in {\it (b)} is $\lambda_z^2 =
y \lambda_x$.
         }
   \label{fig:strupar}
\end{figure}

Fig.  \ref{fig:strupar} displays the spectral distribution of the structure
parameter in the near-wall region and in the outer layer.  Its magnitude is
low around $\lambda_x=\lambda_z$, where it is approximately a function of
the distance to that line, in agreement with the intuitive idea that
isotropic turbulence cannot transport momentum.  On the other hand,
$\sigma_{uv}$ approaches unity for the VLAS, which are thus shown to be
very efficient in transporting momentum.  The result is that they actually
carry an important fraction of the Reynolds stresses in the outer flow, as
shown in Fig.  \ref{fig:uv}.

\section{Discussion and conclusions}

We have performed the first direct numerical simulation of turbulent
channel flow using both a computational domain big enough to capture the
largest structures in the outer flow and a Reynolds number high enough to
observe some separation between those structures and the ones in the
near-wall region.

The results show that there are very large elongated structures in the
outer region of turbulent channel flow whose size scales with $h$.  We have
suggested that they can be understood as the wakes left by compact
isotropic structures decaying under the action of an eddy viscosity as they
are convected by the mean flow.  Both the spectra and flow visualization
suggest that the VLAS are also very high, and that they can hit the walls,
which would help understand the Reynolds number dependence in the scaling
of $\bra u'^2\ket $ in the near-wall region (\PL90; \GE00).  We have seen
that the large structures in the outer layer widen with the wall distance
faster than the buffer layer streaks, reaching widths of order of the
channel height.  The observed widening may be linked to the downstream
evolution of the wakes that we have suggested as the origin of the VLAS.
The large anisotropic structures in the outer flow not only carry a
substantial fraction of the kinetic energy of the flow, but also a
substantial fraction of the Reynolds stresses, and are therefore `active'
in the sense of \cite{Tow76}.

We have noted that our Reynolds number is still too low to draw strong
scaling conclusions for some of the variables involved, because the
separation between the outer and inner scales of the flow is still
moderate.  Computer limitations do not allow direct numerical simulations
in the range of Reynolds numbers in \cite{Hit97}, \cite{KimAdr99},
\cite{MetKle01} and \cite{Moretal01} in a near future, although some of the
open questions can probably be addressed at much lower Reynolds numbers.
Large eddy simulations could be very valuable in this respect if they could
be shown to represent the VLAS sufficiently well.  The wake model that we
have proposed suggests that they should, but it is still not sufficiently
clear what is the origin of the forcing of those wakes, and whether they
are independent of the detailed dynamics of the wall, which is imperfectly
resolved by the LES.\\

This work was supported in part by the Spanish CICYT contract BFM2000-1468
and by ONR grant N0014-00-1-01416.  We are specially indebted to the
CEPBA/IBM center at Barcelona, and to IBM and the U. Polit\`ecnica de
Catalunya, which have gracciously donated the computer time needed for most
of the simulations. Thanks are also due to R.D. Moser who reviewed a
preliminary version of this manuscript.


\end{document}